\documentclass{CRPITStyle} 
\usepackage{amsmath}
\usepackage{amssymb}
\usepackage{url}
\usepackage{graphicx}
\usepackage{calc}
\usepackage[authoryear]{natbib}
\renewcommand{\cite}{\citep}
\def\insimage#1#2#3{
\ifx\pdfoutput\undefined
  \includegraphics[width=#2,height=#3]{#1.eps}
\else
  \pdfximage width #2 height #3 {#1.pdf}
  \mbox{\pdfrefximage \pdflastximage}
\fi}

\def\image#1#2#3#4#5{
\vspace{-2mm}
\begin{figure}[tb]  
\begin{center}
    \ifx\pdfoutput\undefined
       \includegraphics[width=#4,height=#5]{#3.eps}
    \else
      \pdfximage width #4 height #5 {#3.pdf}
      \mbox{\pdfrefximage \pdflastximage}
    \fi      
    \vspace{-2mm}
  \caption{#2}
  \label{#1}
 \end{center} 
\end{figure}}

\def\imagetwoc#1#2#3#4{
\vspace{-2mm}
\begin{figure*}[tb]
\begin{center}
    \ifx\pdfoutput\undefined
       \includegraphics[width=#4]{#3.eps}
    \else
      \pdfximage width #4 {#3.pdf}
      \mbox{\pdfrefximage \pdflastximage}
    \fi      
    \vspace{-2mm}
  \caption{#2}
  \label{#1}
 \end{center} 
\end{figure*}}

\newtheorem{example_th}{Example}
\newtheorem{def_th}{Definition}
\newtheorem{cond_th}{Condition}
\newtheorem{lemma_th}{Lemma}
\newtheorem{listing_th}{Listing}
\newtheorem{theorem_th}{Theorem}

\newenvironment{listing}[3]{\begin{listing_th} \label{#1} \small (#2) \normalfont \smallskip 
\\
\scriptsize
\begin{tabular}{p{8.0cm}}
\hline\\ 
\end{tabular}
{\sffamily \scriptsize #3}\smallskip\\
\begin{tabular}{p{8.0cm}}
\hline\\
\end{tabular}
\end{listing_th}
}{}

\begin{document}

\title{On Metric Skyline Processing by PM-tree}
\author{Tom\'a\v{s} Skopal
\and 
Jakub Loko\v{c}}
\affiliation{Department of Software Engineering, FMP \\
Charles University in Prague, Czech Republic \\
Email:~{\tt skopal@ksi.mff.cuni.cz}\\
Email:~{\tt lokoc@ksi.mff.cuni.cz}}

\maketitle






\begin{abstract}
The task of similarity search in multimedia databases is usually accomplished by range or k nearest neighbor queries. However, the expressing power of these ``single-example'' queries fails when the user's delicate query intent is not available as a single example. Recently, the well-known skyline operator was reused in metric similarity search as a ``multi-example'' query type. When applied on a multi-dimensional database (i.e., on a multi-attribute table), the traditional skyline operator selects all database objects that are not dominated by other objects. The metric skyline query adopts the skyline operator such that the multiple attributes are represented by distances (similarities) to multiple query examples. Hence, we can view the metric skyline as a set of representative database objects which are as similar to all the examples as possible and, simultaneously, are semantically distinct.

In this paper we propose a technique of processing the metric skyline query by use of PM-tree, while we show that our technique significantly outperforms the original M-tree based implementation in both time and space costs. In experiments we also evaluate the partial metric skyline processing, where only a controlled number of skyline objects is retrieved.
\end{abstract}
\vspace{.1in}


\section{Introduction}
As the volumes of complex unstructured data collections grow almost exponentially in time, the attention to content-based similarity search steadily increases. The concept of numeric similarity between two data entities is one of the approaches used in querying unstructured data, where a similarity function serves as multi-valued relevance of data objects to a query (example) object. The content-based similarity search paradigm has been successfully employed in areas like multimedia databases, time series retrieval, bioinformatic and medical databases, data mining, and others. At the same time, the ``similarity-centric'' view on such data demands specific alternative techniques for modeling, indexing and retrieval, which dramatically differ from the traditional approaches to management of structured data (e.g., B-trees in relational databases).

In the rest of the section we introduce into the fundamentals of similarity search and briefly summarize the paper contributions.

\subsection{Similarity search}
Given a collection $\mathcal{C}$ of unstructured data entities (e.g., multimedia objects, like images), to query the collection we need to establish a model consisting of the object \emph{universe} $\mathbb{U}$, a \emph{transformation} function (a feature extraction method, resp.) $t: \mathcal{C} \rightarrow \mathbb{U}$, and a \emph{similarity} function $\delta: \mathbb{U} \times \mathbb{U} \rightarrow \mathcal{R}$. The transformation $t$ turns the collection $\mathcal{C}$ of original data entities into a \emph{database of descriptors} $\mathbb{S} \subset \mathbb{U}$. In most cases the similarity function $\delta$ is expected to be a \emph{metric} distance, because metric properties can be effectively used to index the database $\mathbb{S}$ for efficient (fast) query processing, as discussed later in Section \ref{sec_MAM}.  

\subsubsection{Single-example queries}
The portfolio of available similarity query types consists of mostly single-example queries. The \emph{range} query and \emph{k nearest neighbor} (kNN) query represent the two most popular similarity query types. Using a range query $(Q, r_Q)$ we ask for all objects $O_i \in \mathbb{S}$ the distances of which to a single query object $Q$ are at most $r_Q$. On the other hand, a kNN query $(Q, k)$ selects the $k$ database objects closest to $Q$.

Besides range and kNN queries, there exist some less frequently used query types, like \emph{reverse (k)NN queries}, returning those database objects having the query object $Q$ within their ($k$) nearest neighbor(s). 

\subsubsection{Multi-example queries}
\label{sec_multi}
Although the single-example queries are frequently used nowadays, their expressive power may become unsatisfactory in the future due to increasing complexity and quantity of available data. The acquirement of an example query object is the user's ``ad-hoc'' responsibility. However, when just a single query example should represent the user's delicate intent on the subject of retrieval, finding an appropriate example could be a hard task. Such a scenario is likely to occur when a large data collection is available, and so the query specification has to be fine-grained. Hence, instead of querying by a single example, an easier way for the user could be a specification of several query examples which jointly describe the query intent. Such a multi-example approach allows the user to set the number of query examples and to weigh the contribution of individual examples. Moreover, obtaining multiple examples, where each example corresponds to a partial query intent, is much easier task than finding a single ``holy-grail'' example.

\smallskip
As for existing solutions to multi-example query types, we distinguish three directions. First, there exist many \emph{model-specific techniques} based on an aggregation or unification of the multiple examples, e.g., querying by a centroid in case of vectors, or by a union, intersection or other composition of features of the query examples \cite{TA03}. Although successful in a narrow context (e.g, in image retrieval), this approach is not applicable to the general (metric) similarity case.

Second, a popular approach to multi-query example is issuing multiple single-example queries, while the resulting multiple ranked lists are \emph{aggregated} by means of a \emph{top-k operator} \cite{Fa99}. The advantage of this approach is the employment of an arbitrary aggregation function which provides an important add-on to the expressive power of querying. The main drawback of top-k queries is their high computational and space cost -- there have to be several single-example queries issued and their results materialized and aggregated. 

Third, \emph{browsing} is a retrieval modality where the user plays an important role in the querying process. The user incrementally issues single-example queries, while from the query result of the previous query the user select a new example. Since there are multiple examples used before the user achieves her/his goal, we could interpret browsing as multi-example query processing. The drawback of browsing is obvious, the user is bothered by the enforced interactivity, while the subsequent simple queries may not lead to a satisfactory result anyways.

\smallskip
As other complex operations we name \emph{similarity joins} \cite{JS08}, joining pairs of objects (from one or more databases) based on their proximity, or a special case of the similarity join -- the \emph{closest pair} operator, selecting the two closest objects in the database(s). However, even though the similarity joins provide a complex retrieval functionality, again, they can be regarded as a series of single-example queries, rather than a regular multi-example query type.

\smallskip
In this paper we deal with metric skyline query (detailed in the Section \ref{sec_MSL}), which represents a ``native'' multi-example query type.

\subsection{Metric Access Methods}
\label{sec_MAM}
\begin{figure}[h]

\centering
\includegraphics[width=4cm]{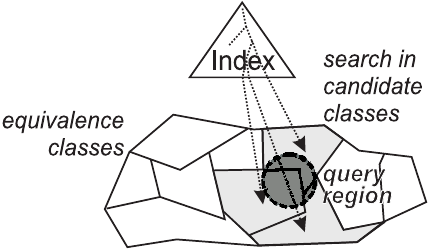}

\caption{Metric access methods.}
\label{img_MAMs}
\end{figure}

When the similarity function $\delta$ is a distance metric, the \emph{metric access methods} (MAMs) can be used for efficient (fast) similarity query processing \cite{Ze+05,Sa06,Ch+01}. The principle behind all MAMs is the utilization of metric postulates (positiveness, symmetry and triangle inequality), which allow to partition the data space into equivalence classes of close (similar) data objects. The classes are embedded within a data structure which is stored in an \emph{index file}, while the index is later used to quickly answer range, kNN, or other similarity queries. In particular, when issued a similarity query, the MAMs exclude many non-relevant equivalence classes from the search (based on metric properties of $\delta$), so only several candidate classes of objects have to be exhaustively (sequentially) searched, see Figure \ref{img_MAMs}. In consequence, searching a small number of candidate classes turns out in reduced costs of the query. 

The number of \emph{distance computations} $\delta(\cdot,\cdot)$ is considered as the major component of the overall costs when indexing or querying a database. Some other cost components (like \emph{I/O costs}, internal \emph{CPU costs}) could be taken into consideration when the computational complexity of $\delta$ is low.

In the following we briefly describe the M-tree and the PM-tree, two MAMs used further in the paper for implementation of metric skyline queries. 

\subsubsection{M-tree}

\begin{figure}[h]
\centering
\includegraphics[width=8.5cm]{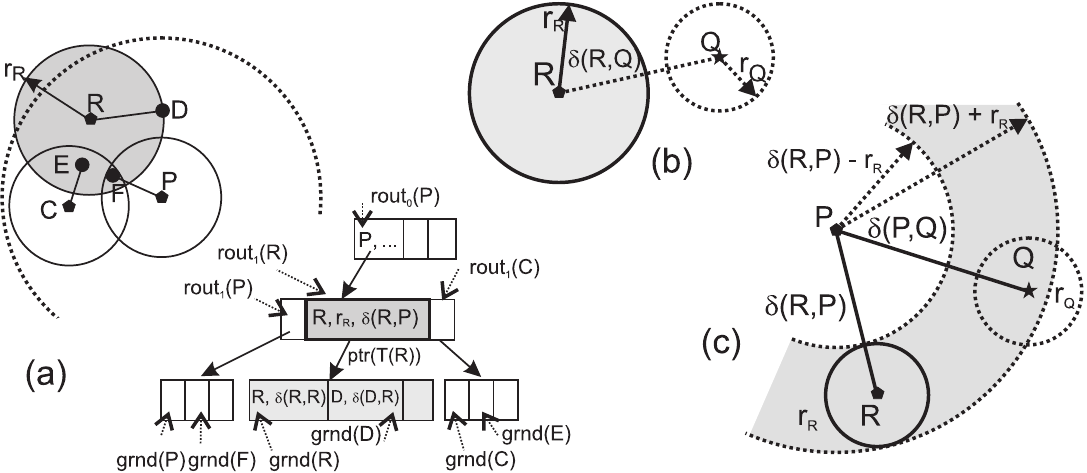}
\caption{(a) M-tree (b) Basic filtering (c) Parent filtering.}
\label{fig_M}
\end{figure}

The \emph{M-tree} \cite{CPZ97} is a dynamic metric access method that provides good performance in database environments. The M-tree index is a hierarchical structure, where some of the data objects are selected as centers (references or local \emph{pivots}) of ball-shaped regions, and the remaining objects are partitioned among the regions in order to build up a balanced and compact hierarchy, see Figure~\ref{fig_M}.

Each region (subtree) is indexed recursively in a B-tree-like (bottom-up) way of construction. The inner nodes of M-tree store \emph{routing entries} 
$$rout_M(R) = [R, r_R, \delta(R, \mathrm{Par}(R)), ptr(T(R))]$$ 
where $R$ is a data object representing the center of the respective ball region, $r_R$ is a \emph{covering radius} of the ball, $\delta(R, Par(R))$ is so-called \emph{to-parent} distance (the distance from $R$ to the object $P$ of the parent routing entry), and finally $ptr(T(R))$ is a pointer to the entry's subtree $T(R)$. In order to correctly bound the data in $T(R)$'s leaves, the routing entry must satisfy the \emph{nesting condition}: $\forall O_i \in T(R), r_R \geq \delta(R, O_i)$. The data is stored in the leaves of M-tree. Each leaf contains \emph{ground entries} 
$$grnd_M(D) = [D, id(D), \delta(D, \mathrm{Par}(D))]$$
where $D$ is the data object itself (externally identified by $id(D)$), and $\delta(D, Par(D))$ is, again, the to-parent distance. See an example of entries in Figure \ref{fig_M}a.

The queries are implemented by traversing the tree, starting from the root. Those nodes are accessed, the parent regions of which are overlapped by the query region, e.g., by a range query ball ($Q, r_Q$). The check for region-and-query overlap requires an explicit distance computation $\delta(R,Q)$ (called \emph{basic filtering}). In particular, if \hbox{$\delta(R,Q) \leq r_Q + r_R$}, the data ball $(R, r_R)$ overlaps the query $(Q, r_Q)$, thus the child node has to be accessed, see Figure \ref{fig_M}b. If not, the respective subtree is filtered from further processing. Moreover, each node in the tree contains the distances from the routing/ground entries to the center of its parent routing entry (the to-parent distances). Hence, some of the M-tree branches can be filtered without the need of a distance computation, thus avoiding the ``more expensive'' basic overlap check. In particular, if \hbox{$|\delta(P,Q) - \delta(P,R)| > r_Q + r_R$}, the data ball $R$ cannot overlap the query ball  (called \emph{parent filtering}), thus the child node has not to be re-checked by basic filtering, see Figure \ref{fig_M}c. Note $\delta(P,Q)$ was already computed at the unsuccessful parent's basic filtering.

\subsubsection{PM-tree}
\label{sec_PMtree}

\begin{figure}[h]
\centering
\includegraphics[width=7.2cm]{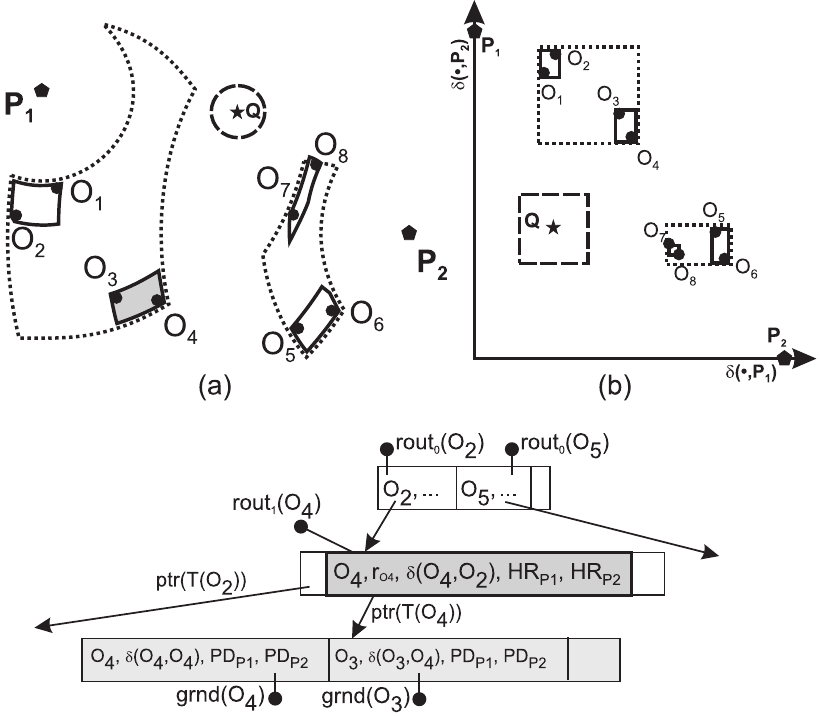}
\caption{(a) PM-tree employing 2 pivots (P$_1$, P$_2$). (b) Projection of PM-tree into the ``pivot space''.}
\label{fig_PM}
\end{figure}

The idea of PM-tree \cite{Sk04,SPS05} is to enhance the hierarchy of \hbox{M-tree} by an information related to a static set of $p$ global pivots $P_i \in \mathcal{P} \subset \mathbb{U}$. In a \hbox{PM-tree's} routing entry, the original M-tree-inherited ball region is further cut off by a set of \emph{rings} (centered in the global pivots), so the region volume becomes more compact -- see Figure~\ref{fig_PM}a. Similarly, the PM-tree ground entries are enhanced by distances to the pivots, which are interpreted as rings as well. Each ring stored in a routing/ground entry represents a distance range (bounding the underlying data) with respect to a particular pivot. 

\medskip
A routing entry in PM-tree inner node is defined as:
$$ rout_{PM}(R) = [R, r_{R}, \delta(R, \mathrm{Par}(R)), ptr(T(R)), \mathrm{HR}],$$
where the new HR attribute is an array of $p_{hr}$ intervals ($p_{hr} \leq p$), where the \hbox{$t$-th} interval HR$_{P_t}$ is the smallest interval covering distances between the pivot $P_t$ and each of the objects stored in leaves of $T(R)$, i.e. HR$_{P_t}=\langle$HR$_{P_t}^{min}$, HR$_{P_t}^{max} \rangle$, HR$_{P_t}^{min} = min\{ \delta(O_j, P_t)\}$,  HR$_{P_t}^{max} = max\{\delta(O_j,P_t)\}$, \hbox{$\forall O_j \in T(R)$}. The interval HR$_{P_t}$ together with pivot $P_t$ define a ring region $(P_t, $HR$_{P_t})$; a ball region $(P_t, $HR$_{P_t}^{max})$ reduced by a "hole" $(P_t, $HR$_{P_t}^{min})$.

\medskip
A ground entry in PM-tree leaf is defined as:
$$ grnd_{PM}(D) = [D, id(D), \delta(D, \mathrm{Par}(D)), \mathrm{PD}],$$ 
where the new PD attribute stands for an array of $p_{pd}$ pivot distances ($p_{pd} \leq p$) where the $t$-th distance PD$_{P_t} = \delta(R, P_t)$.

\medskip
The combination of all the $p$ entry's ranges produces a $p$-dimensional minimum bounding rectangle (MBR), hence, the global pivots actually map the metric regions/data into a ``pivot space'' of dimensionality $p$ (see Figure \ref{fig_PM}b). The number of pivots can be defined separately for routing and ground entries -- we typically choose less pivots for ground entries to reduce storage costs (i.e., $p = p_{hr} > p_{pd}$).

\medskip
When issuing a range or kNN query, the query object is mapped into the pivot space -- this requires $p$ extra distance computations $\delta(Q, P_i), \forall P_i \in \mathcal{P}$. The mapped query ball $(Q, r_Q)$ forms a hyper-cube $\langle \delta(Q, P_1) - r_Q, \delta(Q, P_1) + r_Q \rangle \times \dots \times \langle \delta(Q, P_p) - r_Q, \delta(Q, P_p) + r_Q \rangle$ in the pivot space that is repeatedly utilized to check for an overlap with routing/ground entry's MBRs (see Figures \ref{fig_PM}a,b). If they do not overlap, the entry is filtered out without any distance computation, otherwise, the M-tree's filtering steps (parent \& basic filtering) are applied. Actually, the MBRs overlap check can be also understood as L$_\infty$ filtering, that is, if the L$_\infty$ distance\footnote{The maximum difference of two vectors' coordinate values.} from a PM-tree region to the query object $Q$ is greater than $r_Q$, the region is not overlapped by the query. 

Note the MBRs overlap check does not require an explicit distance computation, so the PM-tree usually achieves significantly lower query costs when compared with M-tree -- up to an order of magnitude (see \cite{Sk04,Sk07,SPS05}).

\subsection{Paper Contributions}
In this paper we introduce metric skyline processing by use of PM-tree, which is a metric access method suitable for similarity search in large databases. We follow the pioneer work \cite{CL08} where the concept of metric skyline query was introduced, and its implementation utilizing M-tree was proposed. In Section \ref{sec_MSL} the metric skyline query and its original implementation is discussed, while in Section \ref{sec_MSLPM} we propose our original PM-tree implementation of metric skyline processing. In experimental results (Section~\ref{sec_exp}) we show that PM-tree based metric skyline processing outperforms the original M-tree implementation not only in terms of distance computation costs, but also in terms of I/O costs, internal CPU costs and internal space costs. 

\section{Metric Skyline Queries}
\label{sec_MSL}
In relational databases, the multi-criterial retrieval is popular in situations where a query exactly specifying the desired attribute ranges cannot be effectively issued. Instead, there is a need for a simplified query concept which selects the desired database objects by some aggregation technique. 

\medskip
Besides the top-k queries \cite{Fa99}, a popular multi-criterial retrieval technique is the \emph{skyline operator} \cite{BKS01}. 

\subsection{The Skyline Operator}
The traditional skyline operator is an advanced retrieval technique that selects objects from a multidimensional database that are ``the best'' from the global point of view. The only assumption on the database is that the attribute domains (dimensions) are linearly ordered, such that the lower (or higher) value of an attribute is, the better the object is (in that attribute). In the rest of the paper we suppose the convention that a lower value in an attribute is better. 

\begin{figure}[h]
\centering
\includegraphics[width=3.5cm]{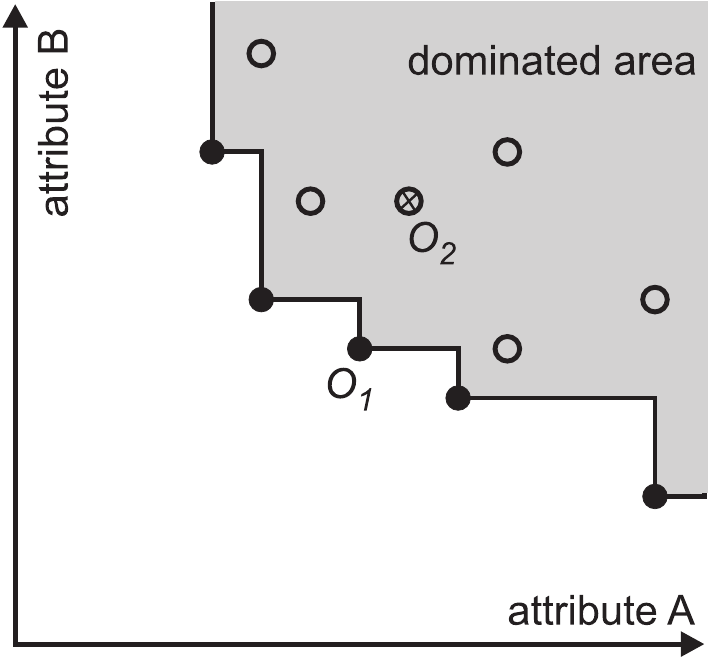}
\caption{A skyline set and the dominated objects.}
\label{fig_skyline}
\end{figure}

\medskip
The skyline operator selects all objects from the database (the \emph{skyline set}), that are not \emph{dominated} by any other object. An object $O_1$ dominates another object $O_2$ if at least one of $O_1$'s attribute values is lower than the same attribute in $O_2$, and the other attribute values in $O_1$ are lower or equal to the corresponding attribute values in $O_2$. Hence, $O_1$ is the \emph{dominating} object, while $O_2$ is the \emph{dominated} object. In Figure \ref{fig_skyline} see an example of skyline set consisting of 5 objects, dominating the remaining 6 objects.

\subsubsection{Skyline Processing}
\label{sec_skylineprocessing}
There exist many approaches to the efficient implementation of the skyline operator, while we outline two of them -- the \emph{Sort-First Skyline} algorithm~\cite{BKS01} and the \emph{branch-and-bound} algorithm which will be useful further in the paper.

In the \emph{Sort-First Skyline} algorithm, the database objects $O_i$ are just ordered ascendentally based on the L$_1$ norm on attributes (coordinates) of $O_i$, i.e., $||O_i||_{L_1} = O_i^1 + O_i^2 + \dots + O_i^n$. Then, following the L$_1$ order, the sorted database is passed such that each visited object $O_i$ is checked whether it is dominated by the already determined skyline objects. If $O_i$ is not dominated, it is added to the skyline set (empty at the beginning), otherwise, $O_i$ is ignored. After the one-pass database traversal is finished, the skyline set is complete.

\begin{figure}[h]
\centering
\includegraphics[width=8cm]{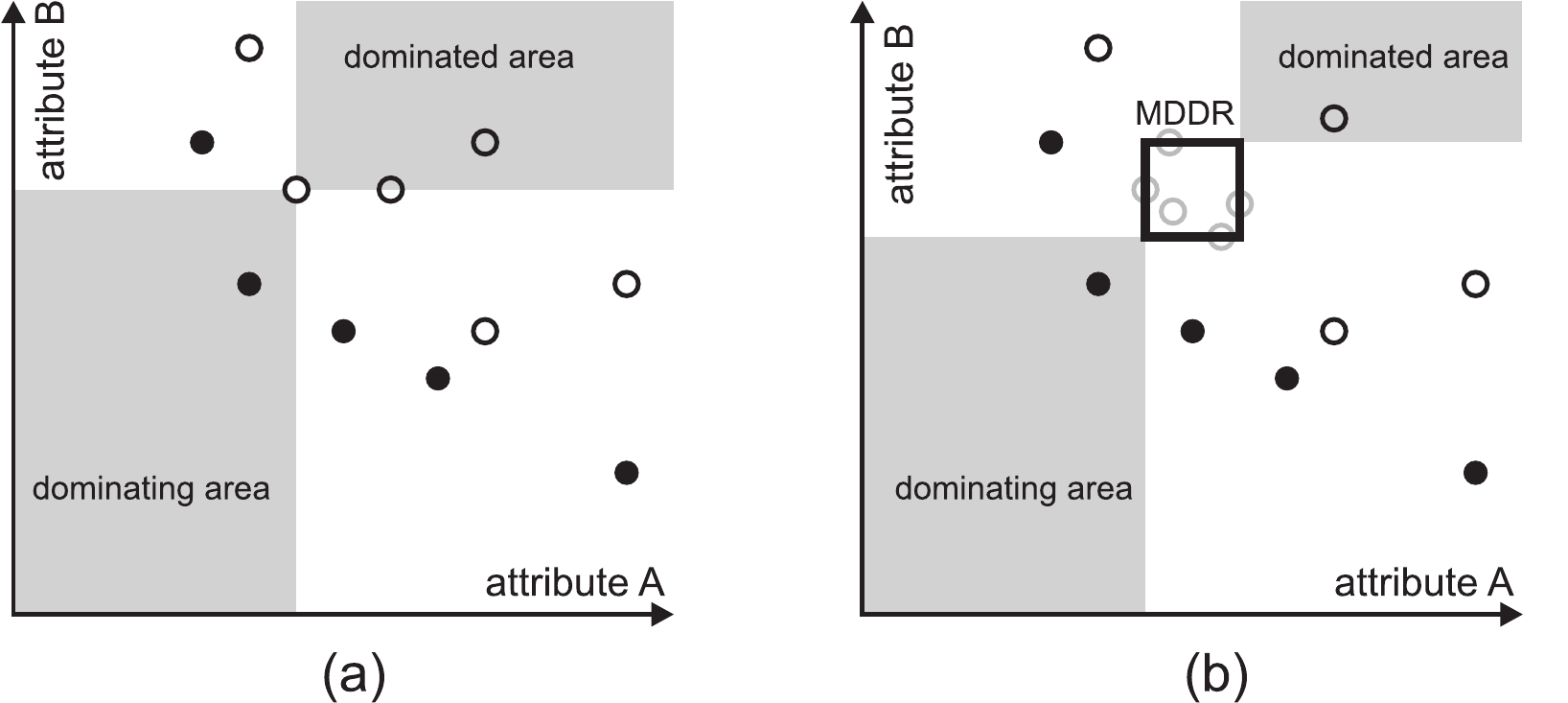}
\caption{A dominating-dominated (a) object and (b) rectangle (MDDR).}
\label{fig_skylineFilt}
\end{figure}

The above algorithm is correct because of the L$_1$-norm ordering. Suppose an object $O_i$ is being processed (see Figure \ref{fig_skylineFilt}a). Because every object possibly dominating $O_i$ lies in the dominating area, its L$_1$ norm must be lower than that of $O_i$. However, such an object has already been visited (and possibly added to the skyline set) because of the ordered database traversal. Thus, $O_i$ can be either safely added to the skyline set or filtered out.

The \emph{branch-and-bound} approach employs a \emph{spatial access method} (SAM), e.g. the R-tree \cite{Gu84}. The database is indexed by the SAM, while for the skyline processing a memory-resident \emph{priority heap} is additionally utilized. The heap priority is defined, again, as the L$_1$ norm, however, besides the database objects themselves, the heap may contain also minimum bounding rectangles (MBRs, natively maintained by, e.g.,  R-tree). For future use outside the scope of SAM, we call MBRs as \emph{minimum dominating-dominated rectangles} (MDDRs). The MDDRs serve as spatial rectangular approximations of the underlying database objects (or nested MDDRs), while they can be effectively used for filtering. The order of an MDDR within the heap is defined by the L$_1$ norm of its minimal corner (the point of MDDR with minimal values in all dimensions), which is the maximal lower bound to L$_1$ norm of any object inside the MDDR.

The skyline processing starts by inserting the top MDDR (within the SAM hierarchy, e.g., R-tree root) into the empty heap. Then, in every step an entry, either an MDDR or a database object, is popped from the heap, while it is checked whether it is dominated by the already determined skyline objects (see Figure \ref{fig_skylineFilt}b). If an MDDR is popped that is not dominated, its descendants in the SAM hierarchy are fetched and inserted into the heap, otherwise the MDDR is removed from further processing.  If an object is popped from the heap that is not dominated, it is added to the skyline set (otherwise filtered out). The correctness of this algorithm is guaranteed by the L$_1$ ordering of the heap.

\subsubsection{Advanced Skyline Queries}
Recently, the concept of skyline operator has been generalized to fit dynamic conditions, where the database object attributes and/or their values are not static. For example, the \emph{dynamic skylines} \cite{Pa+05} consider the attributes as dimension functions. The \emph{spatial skyline queries} \cite{SS06} treat the attribute values as dynamically computed Euclidean distances from a set of query points (multi-point spatial query). The \emph{multi-source skyline queries} \cite{DZS07} are similar, however, instead of Euclidean distances in continuous space the multi-source skyline queries use the shortest-path distances in a graph (in road network, respectively). As another approach, the \emph{reverse skyline queries} \cite{DS07} return the objects whose dynamic skyline contains the original query object (of the reverse skyline query).

\subsection{Metric Skyline Queries}
The spatial skyline queries were generalized recently to support an arbitrary metric distance $\delta$ (i.e., not just Euclidean), constituting thus the \emph{metric skyline queries} (MSQ) \cite{CL08,CL09}. 

Generally speaking, the metric skyline model just adds an abstract transformation step before the usual skyline processing. The step consists of transformation of a database in a metric space into database in $m$-dimensional vector space through a set $\mathcal{Q}$ of $m = |\mathcal{Q}|$ query examples. In the second step, the traditional skyline operator is performed on the transformed database. In particular, a database object $O_i$ in the metric space is transformed into a vector $V_i$, where its $j$-th coordinate is defined as the distance from $j$-th query to $O_i$, i.e., $V_i = \langle \delta(Q_1, O_i), \delta(Q_2, O_i), \dots, \delta(Q_m, O_i) \rangle, Q_j \in~\mathcal{Q}$. 

\subsubsection{Motivation}
The motivation for MSQ can be seen in the insufficient expressive power of range and kNN queries, as mentioned in Section \ref{sec_multi}. Besides the possibility of employing multiple query examples, the metric skyline query has also another unique property, the absence of query extent, i.e., the query is defined just by the set $\mathcal{Q}$. This property could be seen as both advantage and disadvantage. 


The advantage is that metric skyline query returns all distinct objects from the database that are as similar to the query examples as possible. Hence, we obtain all such objects; we are freed from tuning the precision and recall proportion. When issuing range or kNN queries, we have to specify the query extent (i.e., the query radius or the number of nearest neighbors), which could not be as easy as it seems. In particular, the definition a range query radius requires an expert knowledge of the underlying metric distance, otherwise we obtain too small or too large answer set. The kNN query is more user-friendly, however, the precision/recall problem still remains. 

Unfortunately, the disadvantage of MSQ is the skyline set (answer set) size. If $m = |\mathcal{Q}| = 1$ we obtain a regular 1-NN query. However, with increasing $m$ the skyline size usually grows substantially, while a skyline set size exceeding several percent of the database is usually useless for an end-user. Moreover, the skyline set is not uniquely ordered (unlike range or kNN answers), so a reduction of the skyline set cannot be guided by some internal structure of the answer. Thus, to be discriminative enough, the metric skyline query should employ only a few query examples (say, 2--5).

\subsubsection{M-tree Based Implementation}
\label{sec_MtreeSkyline}
The above described straightforward two-step abstraction is not suitable for implementation of MSQ. An explicit transformation of the original database $\mathbb{S}$ into a metric space would require expensive static preprocessing of the database, consisting of $|\mathcal{Q}|\cdot|\mathbb{S}|$ distance computations, extra storage costs, etc. Remember, the main cost component in similarity search by MAMs is the number of distance computations, so any MSQ algorithm should be designed to avoid computing as many distances as possible. 

The authors of metric skyline queries proposed a native MSQ processing by M-tree \cite{CL08,CL09}, where the transformation step is applied only on a part of the database that cannot be skipped during the processing. Basically, the M-tree based metric skyline algorithm was inspired by the traditional skyline processing by R-tree and the priority heap $\mathcal{H}$ under L$_1$ norm (as described in Section \ref{sec_skylineprocessing}).

In the following we have re-formulated the original description in \hbox{\cite{CL08,CL09}} to the more abstract MDDR formalism, due to its easier extensibility to our original contribution in Section \ref{sec_MSLPM}. 

The modification of the traditional R-tree based skyline processing to the metric case resides in an ``on-the-fly'' derivation of MDDRs, which cover the transformed data objects. Instead of ``native'' R-tree MDDRs (MBRs, resp.), we distinguish two types of derived MDDRs in M-tree, as follows: 

\begin{itemize}
\item The \emph{Par-MDDR} (parent MDDR) of a routing/ground entry $entry(R, r_R, \cdots)$\footnote{For a ground entry $r_R = 0$.}, constructed by use of the parent routing entry $rout(P, \cdots)$ as \\MDDR$_{Par} =\langle LB_{Par}^{Q_1}, UB_{Par}^{Q_1} \rangle \times \cdots \times \langle LB_{Par}^{Q_m}, UB_{Par}^{Q_m} \rangle$, where $LB_{Par}^{Q_i}$ is a lower-bound distance from $Q_i$ to the region $(R, r_R)$ (through its parent $P$), while $UB_{Par}^{Q_i}$ is an upper-bound distance from $Q_i$ to $(R, r_R)$. Thus, $LB_{Par}^{Q_i} = max(\delta(Q_i, P) - (\delta(P, R) + r_R), (\delta(P, R) - r_R) - \delta(Q_i, P), 0)$, and \hbox{$UB_{Par}^{Q_i} = \delta(Q_i, P) + \delta(P, R) + r_R$}.

\item  The \emph{B-MDDR} (basic MDDR), constructed directly from a routing/ground entry as MDDR$_{B} = \langle \delta(Q_1, R)~-~r_R,$ $\delta(Q_1, R) + r_R \rangle \times \cdots \times \langle \delta(Q_m, R) - r_R, \delta(Q_m, R) + r_R  \rangle$. As a consequence, B-MDDR of ground entry is a single point.
\end{itemize}

Obviously, we have chosen the terms ``Par-MDDR'' and ``B-MDDR'' due to the analogy with parent- and basic filtering used when processing a range or kNN query in M-tree. The Par-MDDR of a routing/ground entry can be derived without an explicit distance computation; the $\delta(Q_i, P)$ distances were already computed during the top-down M-tree traversal. The derivation of B-MDDR is more expensive, it requires $m$ 
computations of $\delta(R, Q_i), \forall Q_i \in \mathcal{Q}$. 

\begin{figure}[h]
\centering
\includegraphics[width=8.2cm]{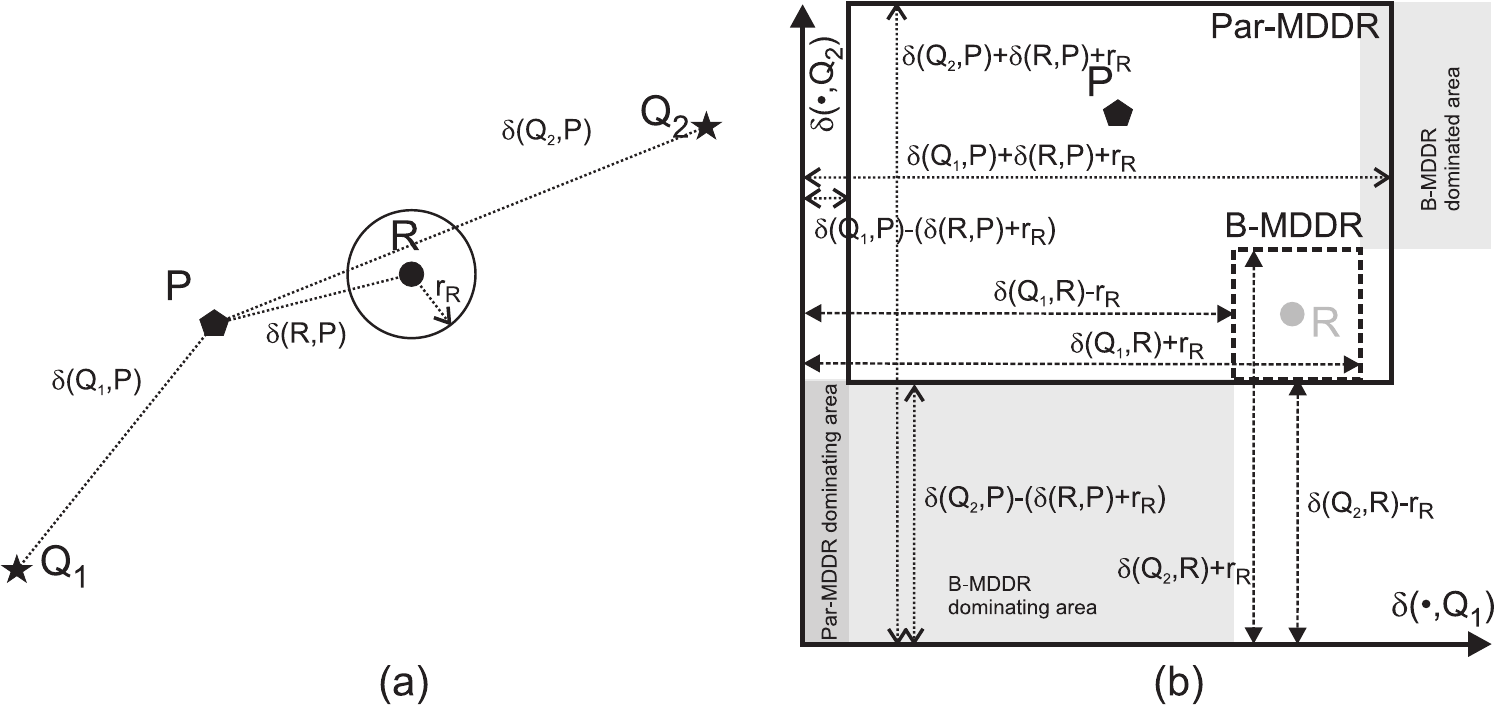}
\caption{(a) Metric space with M-tree regions (b) Transformed vector space with MDDRs}
\label{fig_parentFiltering}
\end{figure}

An MDDR $M_1$ dominates all objects inside an MDDR $M_2$ if the L$_1$ norm of $M_1$'s \emph{maximal corner} is lower than the L$_1$ norm of $M_2$'s \emph{minimal corner}, where a maximal/minimal corner is the point with maximal/minimal values in all dimensions of an MDDR. For an example of Par-MDDR and B-MDDR, see Figure~\ref{fig_parentFiltering}.

The MSQ algorithm starts by inserting routing entries from the M-tree root into the priority heap $\mathcal{H}$. The heap keeps order given by L$_1$ norm applied on the entries' B-MDDRs' minimal corners. Then a loop follows until the heap gets empty:
\begin{enumerate}
\item An entry $entry(R, \dots)$ with the lowest L$_1$ value of its B-MDDR is popped from the heap. 
\item If the entry is a ground entry, it is added to the set of skyline objects. All entries on the heap which are dominated by this new skyline object are removed. Jump to Step~1. 
\item If the entry is a routing entry, the entry's child node is fetched. The Par-MDDRs of the child node's entries are checked for dominance by the set of already determined skyline objects, while the dominated ones (and the respective subtrees, in case of routing entries) are filtered from further processing.
\item The B-MDDRs of the non-filtered child entries are derived. Those entries not dominated by the already retrieved skyline set are inserted into the heap. Jump to Step 1.
\end{enumerate}

\subsubsection{Discussion \& Criticism}
Unfortunately, in the original contribution \cite{CL08,CL09} the cost analysis and also the experiments were focused solely on measuring the number of dominance checks, i.e., how many times B-MDDRs and Par-MDDRs were checked for dominance by a skyline object. The authors completely ignored the number of distance computations (the crucial performance factor for any MAM), but also the heap size and the number of operations on heap, spent by running the metric skyline algorithm on M-tree. 

As we present later in experimental evaluation, the above algorithm, as proposed in \cite{CL08,CL09}, is extremely inefficient in terms of the heap size and the number of operations on the heap. In fact, the maximal heap size could \emph{reach the size of the database} (!), making such an implementation inapplicable in database environments. In the following section we introduce a PM-tree based method, which not only decreases the number of distance computations spent for metric skyline processing, but also drastically decreases the maximal heap size and the number of operations on the heap.

\section{PM-tree based metric skyline}
\label{sec_MSLPM}

The M-tree based approach to metric skyline processing can be extended to a PM-tree based implementation. In the following we introduce an algorithm that makes use of the PM-tree's extensions over the M-tree -- the pivot set $\mathcal{P}$ and the respective ring regions maintained by routing/ground entries in PM-tree nodes (for PM-tree details see Section \ref{sec_PMtree}). 

First of all, when a metric skyline query is started, a \emph{query-to-pivot} matrix of pair-wise distances between the PM-tree pivots $P_i \in \mathcal{P}$ and query examples $Q_i \in \mathcal{Q}$ is computed. The PM-tree based implementation then utilizes the following three filtering concepts (Sections \ref{sec_PivMDDR}--\ref{sec_Incre}), summarized within an algorithm in Section \ref{sec_algPM}:

\subsection{Deriving Piv-MDDRs}
\label{sec_PivMDDR}
Besides the M-tree's B-MDDRs and Par-MDDRs derived from a routing/ground $entry(R, \cdots, \mathrm{HR}/\mathrm{PD})$, an additional MDDR can be derived from the set of rings HR/PD maintained by the entry, called \emph{Piv-MDDR} (pivot MDDR). The Piv-MDDR can be derived using the query-to-pivot matrix, as MDDR$_{Piv} = \langle LB_{Piv}^{Q_1}, UB_{Piv}^{Q_1} \rangle \times \cdots \times \langle LB_{Piv}^{Q_m}, UB_{Piv}^{Q_m} \rangle$, where \\$\mathrm{LB}^{Q_i}_{Piv} = max_{P_j \in \mathcal{P}} \{ \delta(P_j, Q_i) - \mathrm{HR}_{P_j}^{max}, \mathrm{HR}_{P_j}^{min} - \delta(P_j,Q_i), 0 \}$, and
$\mathrm{UB}^{Q_i}_{Piv} = min_{P_j \in \mathcal{P}} \{ \delta(P_j, Q_i) + \mathrm{HR}_{P_j}^{max}\}$.

\begin{figure}[h]
\centering
\includegraphics[width=8.5cm]{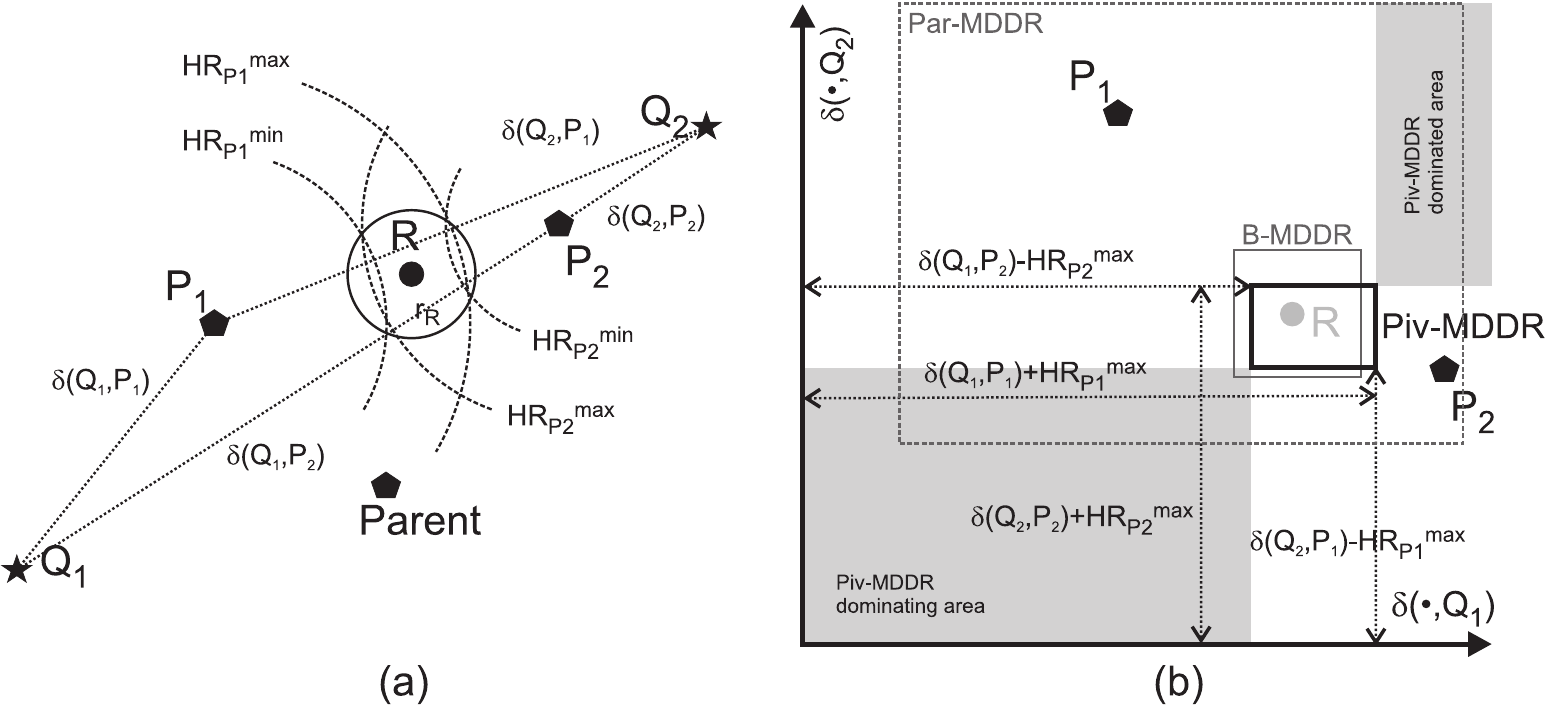}
\caption{A PM-tree routing entry in (a) metric space and (b) mapped to Piv-, Par-, and B-MDDR.}
\label{fig_pivotFiltering}
\end{figure}

\medskip
Similarly as the M-tree's Par-MDDR, the derivation of Piv-MDDR requires no extra distance computation, however, Piv-MDDRs are much more compact than Par-MDDRs. This results in more effective filtering of routing/ground entries by skyline objects or some dominating MDDRs. Moreover, the Piv-MDDR is often even more compact than the direct \hbox{B-MDDR}, because the PM-tree's rings reduce the volume of the original M-tree's sphere. In Figure \ref{fig_pivotFiltering} see an example of Piv-MDDR, Par-MDDR and B-MDDR, when 2-pivot PM-tree and 2 query examples are used.

Naturally, when having two or three $x$-MDDRs available, e.g., B-MDDR + Piv-MDDR, we can intersect them to form a single compact MDDR which is then used for filtering.

\subsection{Pivot-Skyline Filtering}
\label{sec_PSF}

\begin{figure}[h]
\centering
\includegraphics[width=3.5cm]{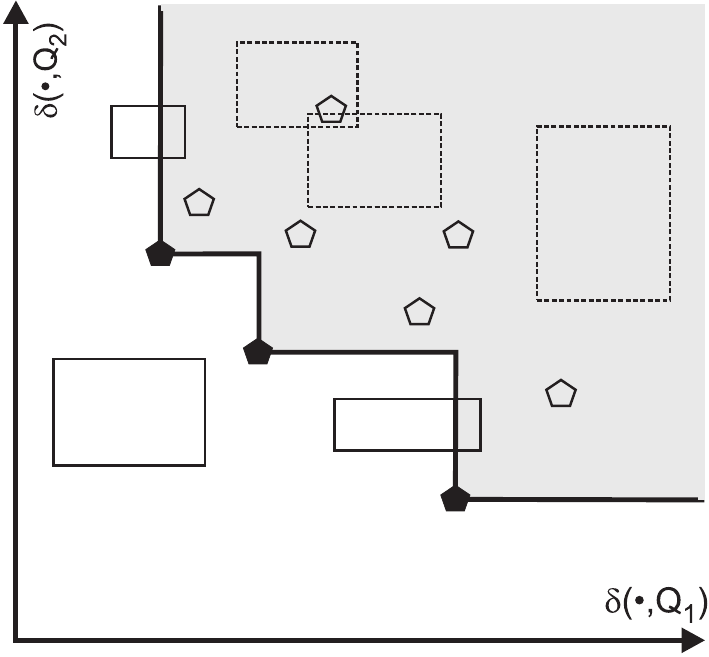}
\caption{A pivot skyline.}
\label{fig_pivotSkyline}
\end{figure}

If the pivots $P_i$ come from the  database (i.e., $P_i \in \mathcal{P} \subset \mathbb{S}$), the MDDRs that are about to be inserted into the heap can be checked for a dominance by the pivots. Since the query-to-pivot matrix is computed at the beginning of every metric skyline query processing, the transformation of the pivots into the ``query space'' requires no additional distance computations. Moreover, to reduce the number of pivots used for dominance checking, we can determine the so-called \emph{pivot skyline} -- those pivot objects, which constitute a metric skyline within the pivot set $\mathcal{P}$ itself, see an example in Figure~\ref{fig_pivotSkyline}. 

The filtering by use of pivot skyline is beneficial in the early phase of the metric skyline processing, when the set of determined skyline objects is still empty. In the experiments we show that such an early phase is the dominant phase of the entire skyline processing -- 80-90\% of the total distance computations is performed before the first skyline object is found. Hence, pruning the heap by use of the pivot skyline greatly helps to reduce the heap size and, consequently, the number of operations on the heap.

\subsubsection{Merging Pivot Skyline with the Regular Skyline}
As the number of determined skyline objects grows, the objects in the pivot skyline become dominated by the ``regular'' skyline objects. Hence, in order to effectively use the pivots for dominance checking, we keep just those pivots in the pivot skyline, that are not dominated by the already determined skyline objects. Thus, at the moment when all skyline objects are known, the pivot skyline becomes empty.

\subsection{Deferred Heap Processing}
\label{sec_Incre}
In the original M-tree algorithm, the priority heap contains just L$_1$-ordered B-MDDRs (together with the associated routing/ground entries). When an entry is to be inserted into the heap, its B-MDDR must be determined, see Steps 3,4 of the algorithm in Section \ref{sec_MtreeSkyline}. We call this approach a \emph{non-deferred heap processing}.

However, the non-deferred heap processing is not optimal in terms of the number of distance computations. In order to save some distance computations, we propose the \emph{deferred heap processing} for the metric skyline, inspired by the Hjaltason's \& Samet's incremental nearest neighbor algorithm, which is optimal in the number of distance computations \cite{HS00}. 

The modified heap is generalized such that it may contain not only B-MDDRs of routing/ground entries, but also the intersections of their Piv-MDDR and Par-MDDR (denoted as Piv-MDDR $\cap $ Par-MDDR).  The deferred heap processing then deals with two situations:
\begin{itemize}
\item An entry equipped by B-MDDR is popped from the heap. Then, \\
(a) If the entry is a ground entry, it becomes a skyline object.\\
(b) If the entry is a routing entry, its child node is fetched, while for every entry in the child node the Piv-MDDR $\cap$ Par-MDDR is checked for a dominance by the skyline set. Every not-dominated child entry is equipped by its Piv-MDDR $\cap$ Par-MDDR and inserted into the heap. 
\item An entry equipped by Piv-MDDR $\cap$ Par-MDDR is popped from the heap. The entry is checked for a dominance by the skyline set. If not dominated, the entry's B-MDDR is determined and, if still not dominated, inserted back into the heap.
\end{itemize}

{\bf Note: } The deferred heap processing ``gives the algorithm a chance'' to filter out as many entries as possible, without the need of B-MDDR derivation (requiring explicit distance computations). On the other hand, the ``reinsertions'' of PM-tree entries into the heap (first, equipped by Piv-MDDR $\cap$ Par-MDDR, and second, equipped by B-MDDR) increase the number of operations on the heap and also the heap size. 

\subsection{The Algorithm}
\label{sec_algPM}

In Listing \ref{list_MSQuery} the algorithm for metric skyline query is presented, including the original M-tree variant as well as the proposed PM-tree extensions. 

The input attribute {\sffamily type} allows to set the MSQ variant as follows: {\sffamily type = 'M-tree'} is the original M-tree based algorithm, {\sffamily type = 'PM-tree'} is the basic PM-tree based algorithm using the Piv-MDDR filtering (as described in Section \ref{sec_PivMDDR}), {\sffamily type = 'PM-tree+PSF'} additionally uses the pivot-skyline filtering (as described in Section \ref{sec_PSF}), and {\sffamily type = 'PM-tree+PSF+DEF'} additionally uses the deferred heap processing (as described in Section \ref{sec_Incre}).

\noindent
\listing{list_MSQuery}{Algorithm of metric skyline query}{

\noindent
{\bfseries MSQuery}()\\
\{\\
{\bfseries Input}:  PM-tree $\mathcal{PM}$, query points $\mathcal{Q}$, type ('M-tree', 'PM-tree',\\ \hglue 1mm \hfill 'PM-tree+PSF', 'PM-tree+PSF+DEF')\\
{\bfseries Output}: Result $\mathcal{MSS}$ containing skyline points\\

\noindent
\hglue 2mm {\bfseries if} (type {\bfseries is not} 'M-tree')\\
\hglue 4mm P2Q\_DM = evaluate the query-to-pivot matrix\\
\hglue 4mm \emph{// pivots must be DB objects}\\
\hglue 4mm PSL = evaluate pivot skyline (using P2Q\_DM)\\

\noindent
\hglue 2mm Insert all routing entries + their Piv-MDDR $\cap$ B-MDDR from the\\ \hglue 1mm \hfill{PM-tree root into the heap $\mathcal{H}$}\\

\noindent
\hglue 2mm {\bfseries while} ($\mathcal{H}$ {\bfseries is not} empty)\\
\hglue 4mm currentEntry = pop entry from the heap $\mathcal{H}$\\

\noindent
\hglue 4mm {\bfseries if} (currentEntry {\bfseries is not equipped by} 'B-MDDR')\\
\hglue 6mm FilterAndInsert(currentEntry, currentEntry, type, {\bfseries true})\\

\noindent
\hglue 4mm {\bfseries else if} (currentEntry {\bfseries is of type} 'ground entry'  {\bfseries and is equipped by}\\ \hglue 1mm \hfill 'B-MDDR')\\
\hglue 6mm Insert currentEntry into $\mathcal{MSS}$\\
\hglue 6mm $\mathcal{H}$.FilterDominatedObjectsBy(currentEntry.MDDR)\\
\hglue 6mm PSL.FilterDominatedObjectsBy(currentEntry.MDDR)\\

\noindent    
\hglue 4mm {\bfseries else}\\
\hglue 6mm $N$ = fetch child node of currentEntry\\
\hglue 6mm {\bfseries for each} childEntry {\bfseries in} $N$\\
\hglue 8mm FilterAndInsert(childEntry, currentEntry, type, {\bfseries false})\\
\}\\


\smallskip

\noindent
{\bfseries FilterAndInsert}(newEntry, parentEntry, type, deferred)\\
\{\\
\hglue 2mm {\bfseries if} ({\bfseries not} deferred)\\
\hglue 4mm Equip newEntry by its Par-MDDR\\
\hglue 4mm {\bfseries if} (type {\bfseries is not} 'M-tree')\\
\hglue 6mm Update newEntry.MDDR by intersection with newEntry's Piv-MDDR\\

\noindent    
\hglue 2mm {\bfseries if} (Filter(newEntry, type))\\
\hglue 4mm {\bfseries return}\\

\noindent
\hglue 2mm {\bfseries if} (type = 'PM-tree+PSF+DEF' {\bfseries and not} deferred)\\
\hglue 4mm Insert newEntry into $\mathcal{H}$\\
\hglue 4mm {\bfseries return}\\

\noindent
\hglue 2mm Equip newEntry by its B-MDDR\\

\noindent
\hglue 2mm {\bfseries if} (Filter(newEntry, type))\\
\hglue 4mm {\bfseries return}\\

\noindent
\hglue 2mm Insert newEntry into $\mathcal{H}$\\
\}\\

\pagebreak

\noindent
{\bfseries Filter}(newEntry, type)\\
\{\\
\hglue 2mm {\bfseries for each} $O_i$ {\bfseries in} $\mathcal{MSS}$\\
\hglue 4mm {\bfseries if} (newEntry.MDDR is dominated by $O_i$)\\
\hglue 6mm {\bfseries return true}\\

\noindent
\hglue 2mm {\bfseries if} (type {\bfseries is} 'M-tree' {\bfseries or} 'PM-tree')\\
\hglue 4mm {\bfseries return false}\\

\noindent
\hglue 2mm {\bfseries for each} $O_i$ {\bfseries in} PSL\\
\hglue 4mm {\bfseries if} (newEntry.MDDR is dominated by $O_i$)\\
\hglue 6mm {\bfseries return true}\\

\noindent
\hglue 2mm {\bfseries return false}\\
\}
}

\subsection{Runtime Properties}
Since the thread of metric skyline processing may generally follow many different scenarios (depending on the data distribution, metric distance employed, number of pivots, number of queries, etc.), the time and space costs cannot be exactly determined beforehand. Nevertheless, we could observe some properties that will (more or less) occur for any set of conditions. 

First, the algorithm of metric skyline query uses the priority heap (either the non-deferred or deferred variant) and second, an object already inserted into the skyline set remains a skyline object forever. The second observation gives a clue to the typical heap evolution. Because an object is inserted into the skyline set after it is definitely clear it belongs to the skyline, one can conclude that a large proportion of the entire query logic must be performed before the first skyline object is reached. We call this early phase an \emph{expansion phase}, because the heap content cannot be pruned by the dominating skyline objects (they do not exist yet), and so the heap size only grows (expands). After the skyline set begins to populate, the heap begins to shrink, because the insertions of child routing/ground entries into the heap are compensated by removals of the dominated MDDRs. We call the second phase a \emph{reduction phase}.

The expansion phase can be shortened by a utilization of the pivot-skyline filtering, however, the impact of the regular skyline objects is much greater -- they dominate much more objects/MDDRs due to their lower L$_1$ distances.

As we show in the experimental evaluation, the expansion phase takes 80-90\% of the time, when measuring the time as the proportion of distances computed so far to the total number of distance computations. When measuring the time in terms of heap operations, the expansion phase takes 25-75\% of the time, while the heap size is maximal right before the reduction phase begins.

\subsubsection{Partial Metric Skyline}
\label{sec_partial}
As the set of already determined skyline objects can only grow, it is easy to adopt the metric skyline algorithm to provide \emph{partial metric skyline queries}, where the user specifies only a limited number of skyline objects she/he wants to retrieve. The algorithm simply terminates as soon as the specified number of skyline objects appears in the skyline set.

Unfortunately, due to the runtime properties described above, the performance of partial metric skyline query does not scale well with the number of desired skyline objects. Even a single retrieved skyline object requires 25-75\% of heap operations and 80-90\% of distance computations, when compared to the ``full'' metric skyline query.

\section{Experimental Evaluation}
\label{sec_exp}
We performed an extensive experimentation with the three new variants of the PM-tree based metric skyline processing, comparing them against the original \hbox{M-tree} based method. Instead of the number of dominance checks (as included in the original contribution \cite{CL08,CL09}), we have observed other 4 measures of costs spent by the MSQ processing -- the number of distance computations, the number of operations on the heap, the maximal allocated size of the heap, and finally the I/O costs. 

In addition to the absolute numbers presented in the figures below, we also relate the number of distance computations spent by (P)M-tree MSQ processing to the costs of MSQ processed by simple \emph{sequential search}, which takes $|\mathcal{Q}| \cdot |\mathbb{S}|$ distance computations for every query.

\subsection{The Testbed}

We have used two databases, a subset of the \emph{CoPhIR} database \cite{cophir} of MPEG7 image features extracted from images downloaded from {\tt flickr.com}, and a synthetic database of polygons. The CoPhIR database, consisting of one million feature vectors, was projected into two subdatabases, the \emph{CoPhIR\_12} database, consisting of 12-dimensional color layout descriptors, and the \emph{CoPhIR\_76} database, consisting of 76-dimensional descriptors (12-dimensional color layout and 64-dimensional color structure). As a distance function the Euclidean ($L_2$) distance was employed. 

The \emph{Polygons} database was a synthetic randomly generated set of 250,000 2D polygons, each polygon consisting of 5--15 vertices. The Polygons should serve as a non-vectorial analogy to clustered points. The first vertex of a polygon was generated at random. The next one was generated randomly, but the distance from the preceding vertex was limited to 10\% of max. distance in the space. We used the Hausdorff distance \cite{HKR93} for measuring the distance between two polygons, so here a polygon could be interpreted as a cloud of points.

\subsection{Experiment Settings}

The query costs were always averaged for 200 metric skyline queries, while the query examples followed the distribution of database objects. As the parameters we observed various database sizes, the (P)M-tree node capacities, the number of query examples, the size of partial metric skyline, and the number of PM-tree leaf pivots. The (P)M-tree node capacities ranged from 20 to 40 routing/ground entries, the index sizes took 200MB--2GB, the P(M)-tree heights were 3--5 (4--6 levels). The minimal (P)M-tree node utilization was set to 20\% of node capacity. The number of PM-tree leaf pivots ranged from 30 to 1000, while the number of inner pivots ranged from 15 to 500. Unless otherwise stated, the number of MSQ query examples was 2, the (P)M-tree node size was 20, the number of leaf pivots was 1000 for CoPhIR and 300 for Polygons (the number of inner pivots was half the number of leaf pivots). 

\subsection{The Results}
\begin{figure}[h]
\centering
\includegraphics[width=3.9cm]{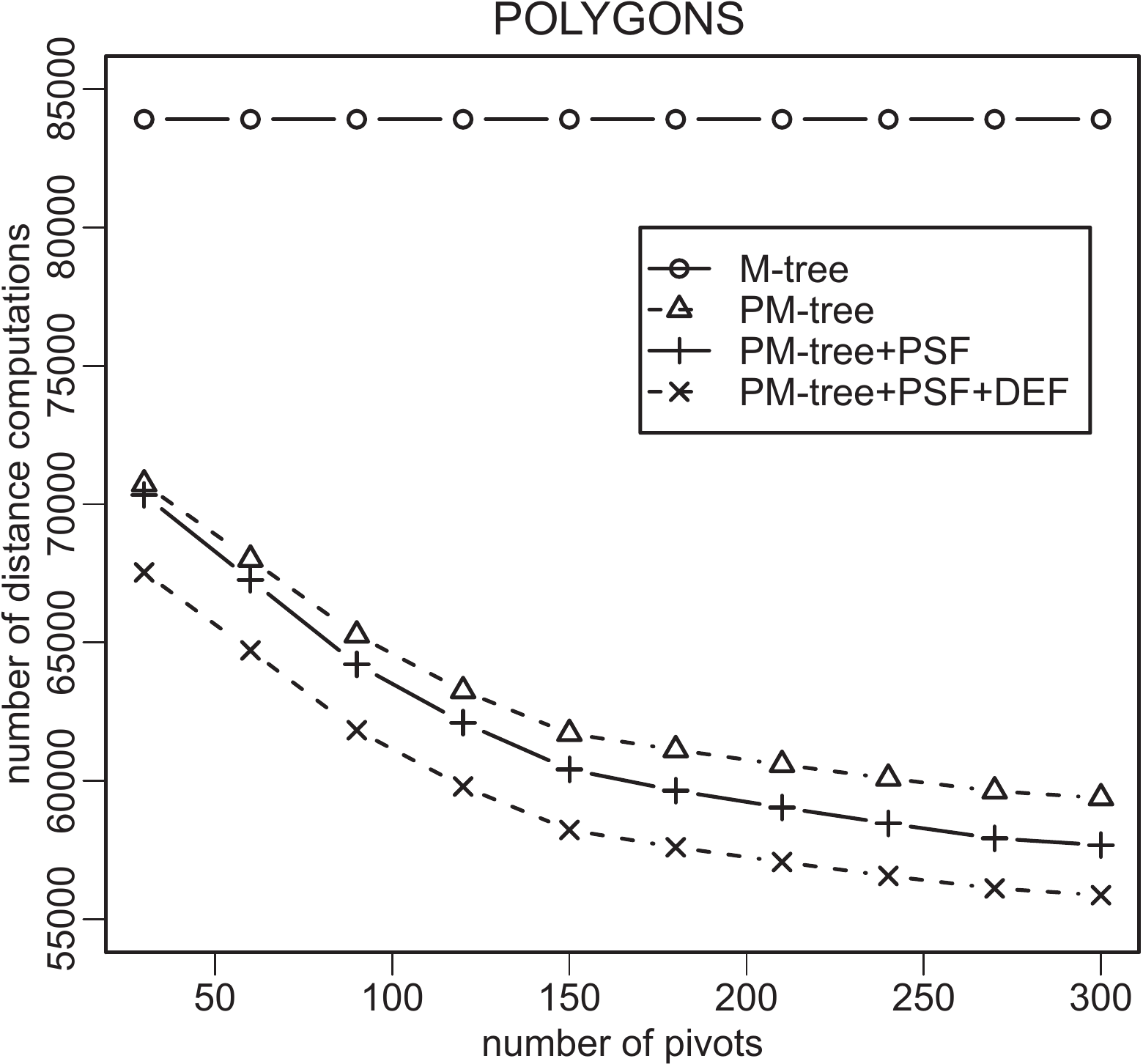}
\includegraphics[width=3.9cm]{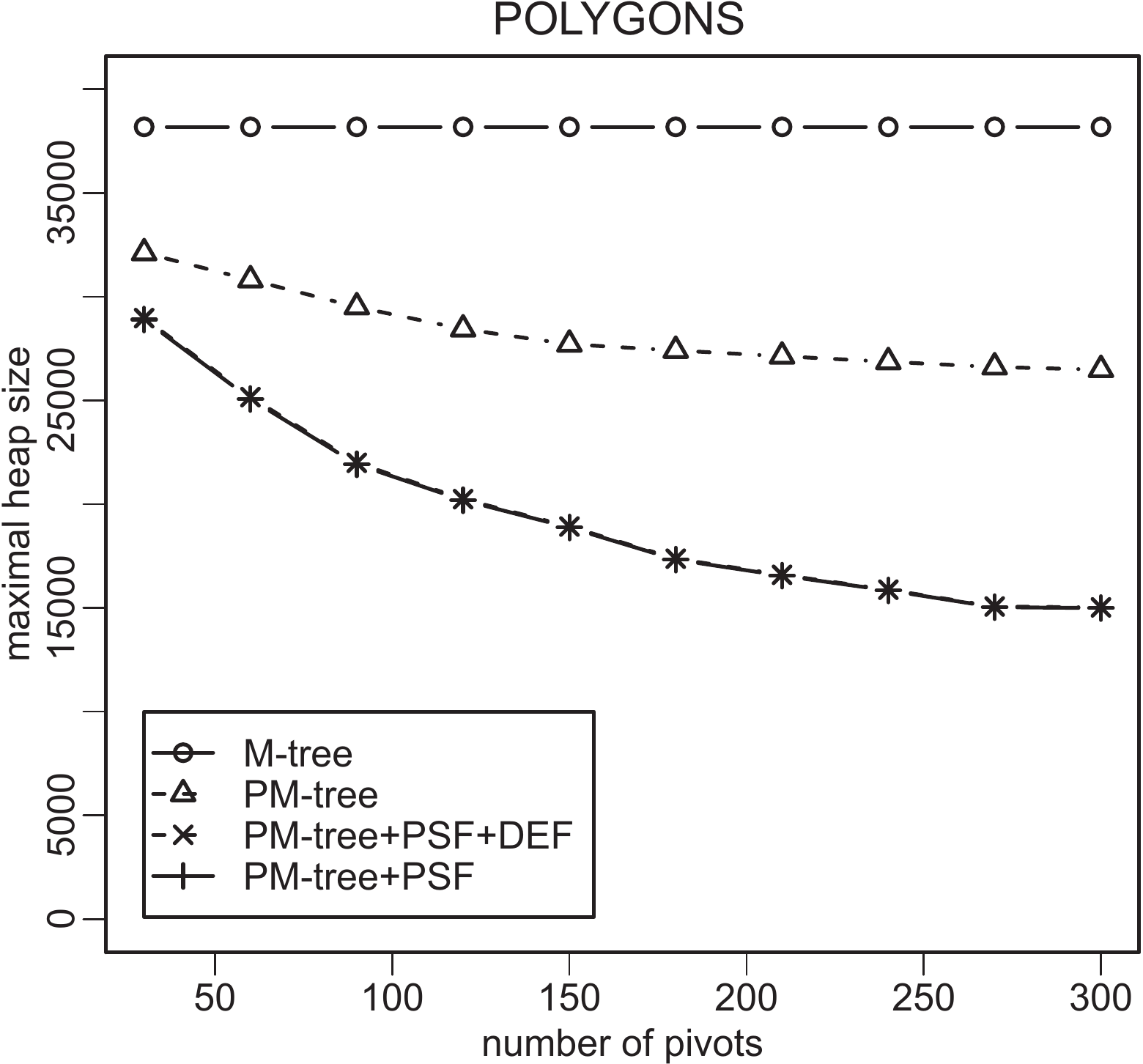}
\caption{Increasing number of pivots on Polygons: (a) Distance computations (b) Maximal heap size }
\label{fig_Exp1}
\end{figure}

In the first set of experiments, the number of PM-tree leaf pivots was increasing. In Figure \ref{fig_Exp1}a the M-tree's MSQ got to 17\% of distance computations needed by simple sequential search on the Polygons database. However, for the highest number of pivots the PM-tree's MSQ reduced the M-tree costs by another 35\%. The heap size required by PM-tree reached only up to one third of the heap size required by the M-tree (see Figure \ref{fig_Exp1}b). The impact of pivot-skyline filtering (the +PSF(+DEF) variants) on the maximal heap size was significant.

\begin{figure}[h]
\centering
\includegraphics[width=3.9cm]{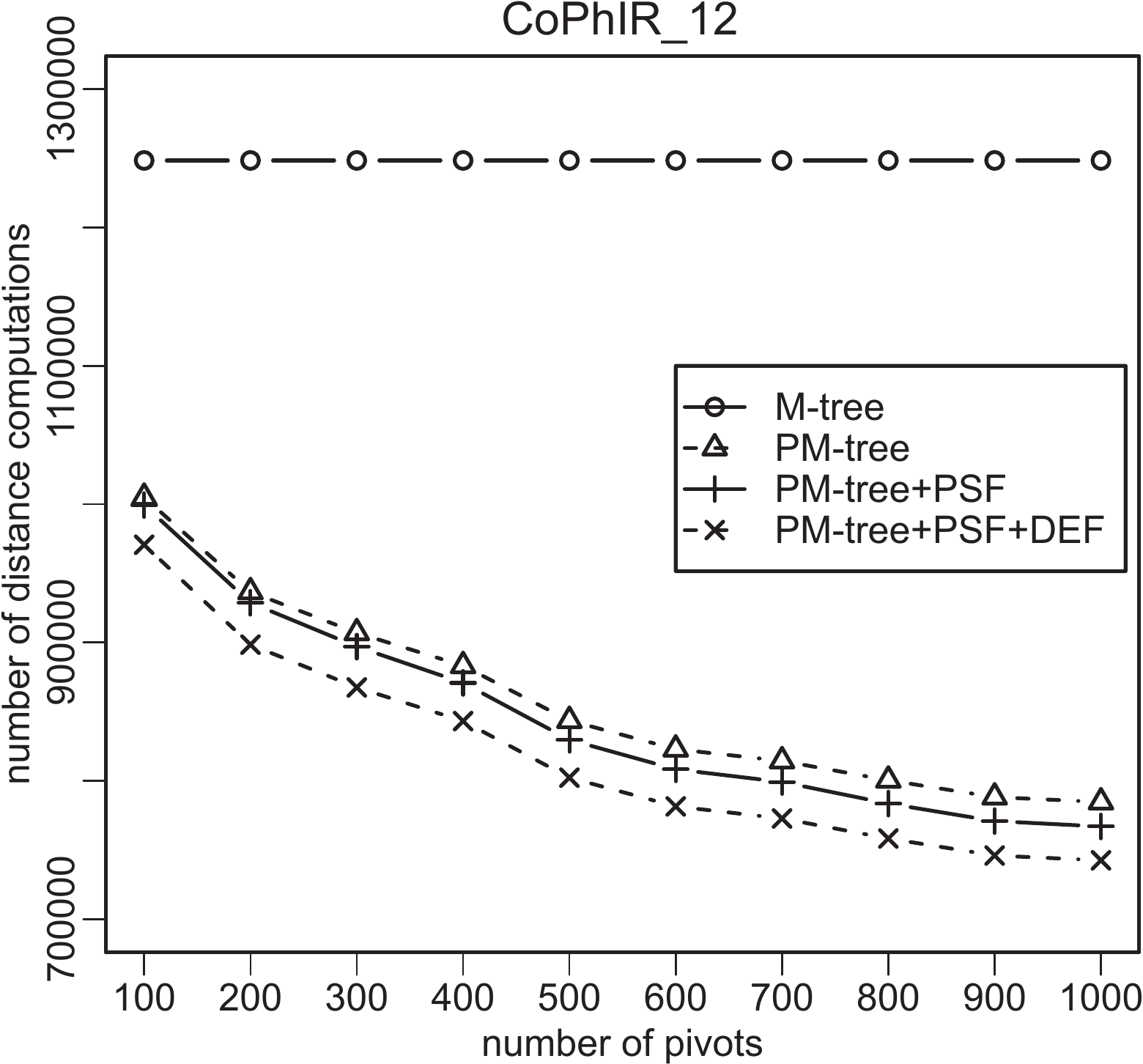}
\includegraphics[width=3.9cm]{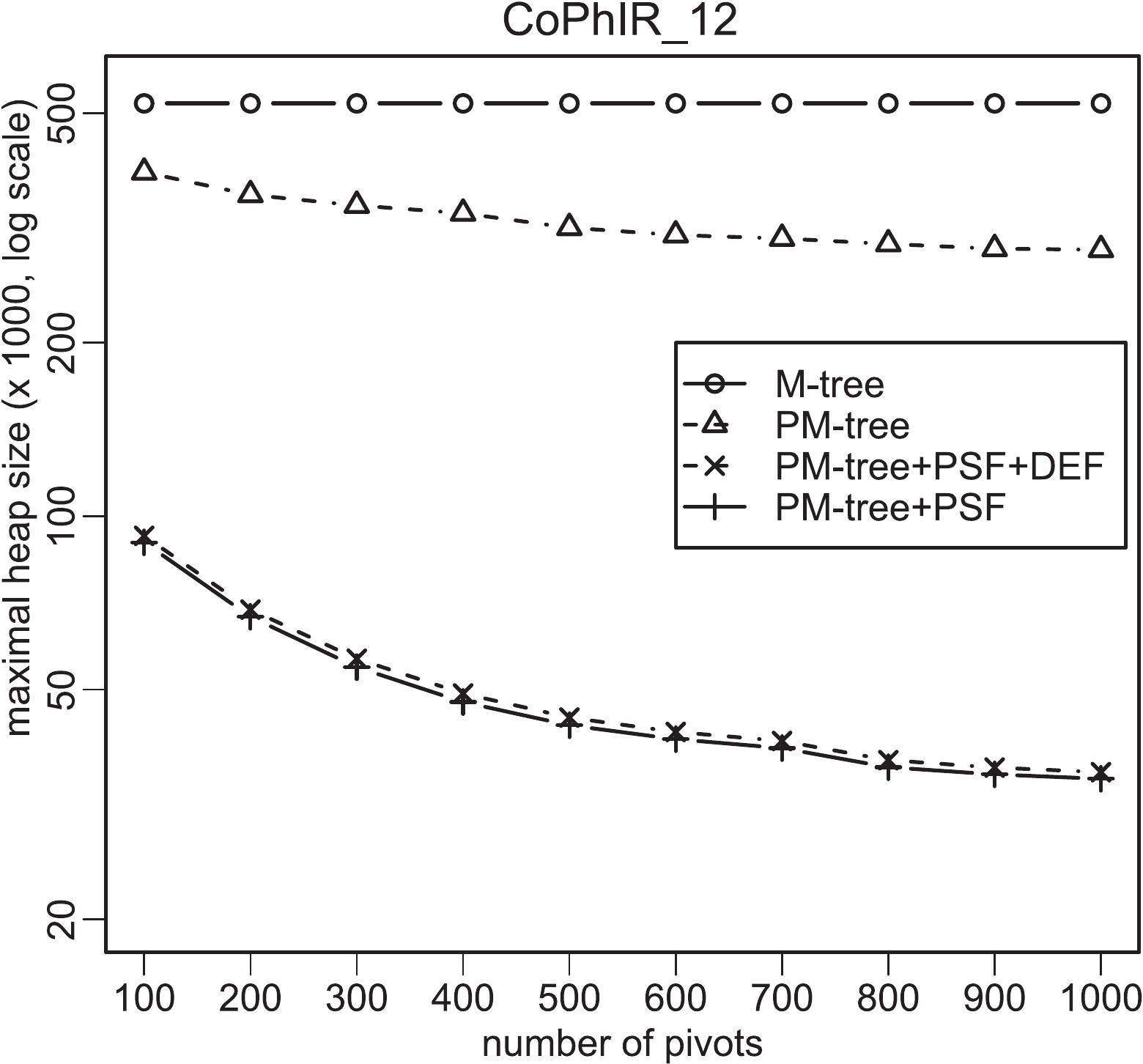}
\caption{\hbox{Increasing number of pivots on Cophir\_12:} (a) Distance computations (b) Maximal heap size }
\label{fig_Exp2}
\end{figure}

In Figure \ref{fig_Exp2} the same situation is presented for the Cophir\_12 database. The results are even better as for Polygons -- the number of distance computations for PM-tree+PSF+DEF variant was reduced to 60\% of M-tree costs, while the maximal heap size was reduced down to 8\% of the heap size required by M-tree (note the log.scale in Figure \ref{fig_Exp2}b).

\begin{figure}[h]
\centering
\includegraphics[width=3.9cm]{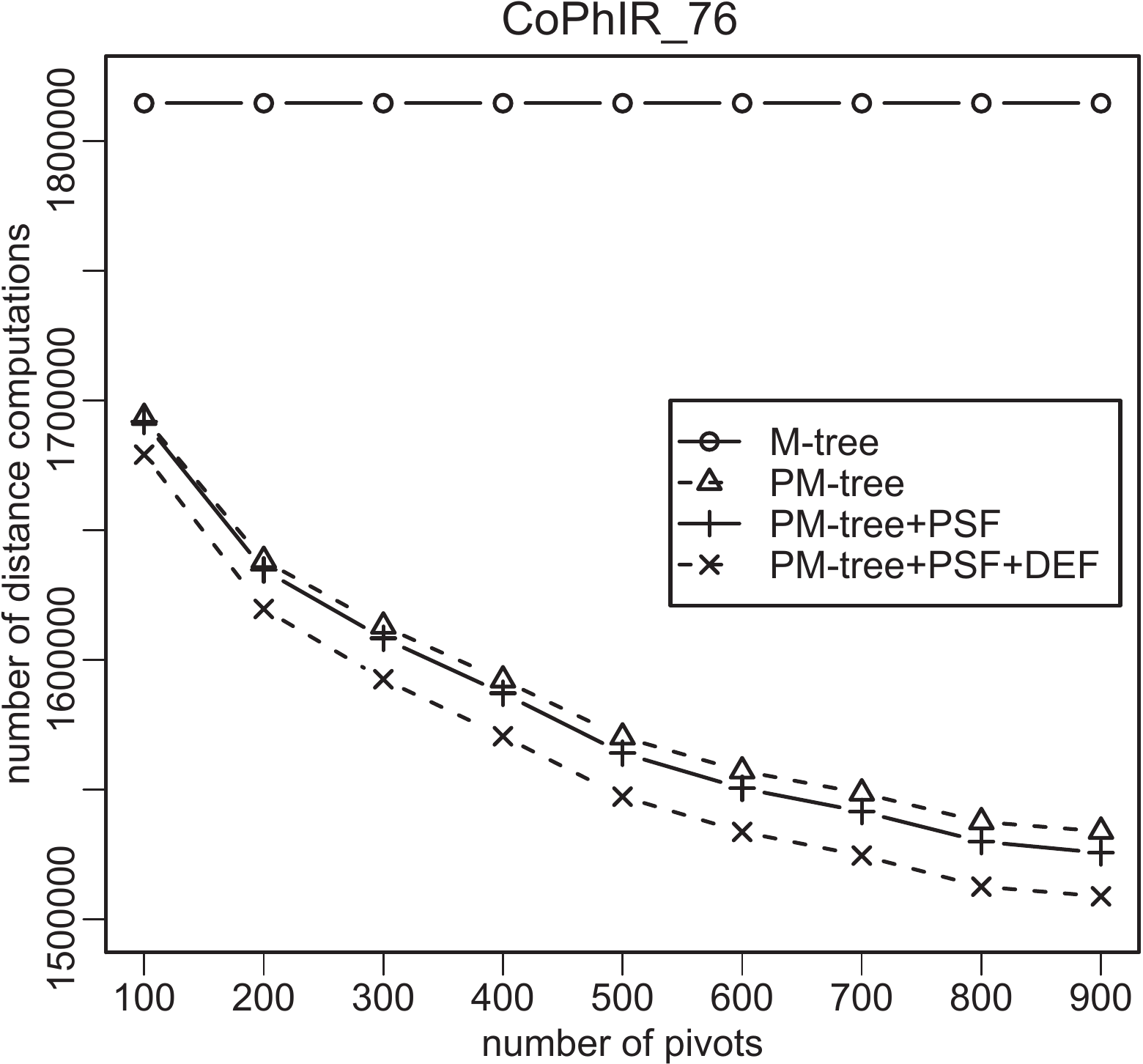}
\includegraphics[width=3.9cm]{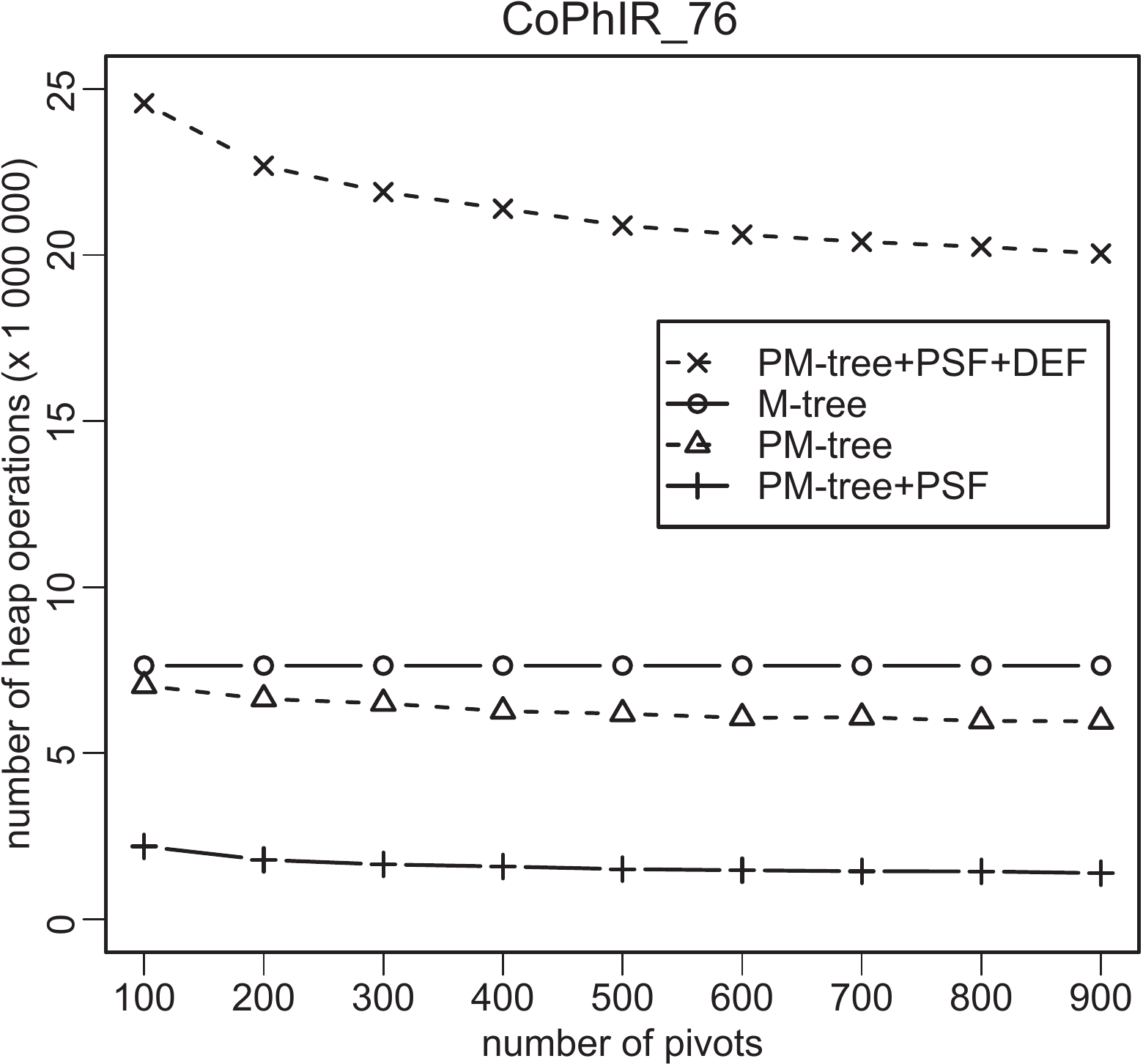}
\caption{\hbox{Increasing number of pivots on Cophir\_76:} (a) Distance computations (b) Heap operations}
\label{fig_Exp3}
\end{figure}

Finally, in Figure \ref{fig_Exp3} the same situation is presented for the high-dimensional Cophir\_76 database. Because of the high dimensionality, the M-tree performance was poor -- it got to 91\% distance computations required by simple sequential search (see Figure \ref{fig_Exp3}a). The PM-tree performed better, achieving 75\% of the sequential search's distance computations.
In Figure \ref{fig_Exp3}b the number of heap operations is presented. Here the PM-tree+PSF+DEF variant performs poorly, because of the deferred heap processing, i.e., repeated insertions of MDDRs into the heap (see Section \ref{sec_Incre}). On the other hand, the +DEF variant steadily achieves the lowest distance computation costs (as expected). The PM-tree+PSF variant performs the best, achieving 25\% of the heap operations spent by M-tree.

\begin{figure}[h]
\centering
\includegraphics[width=4cm]{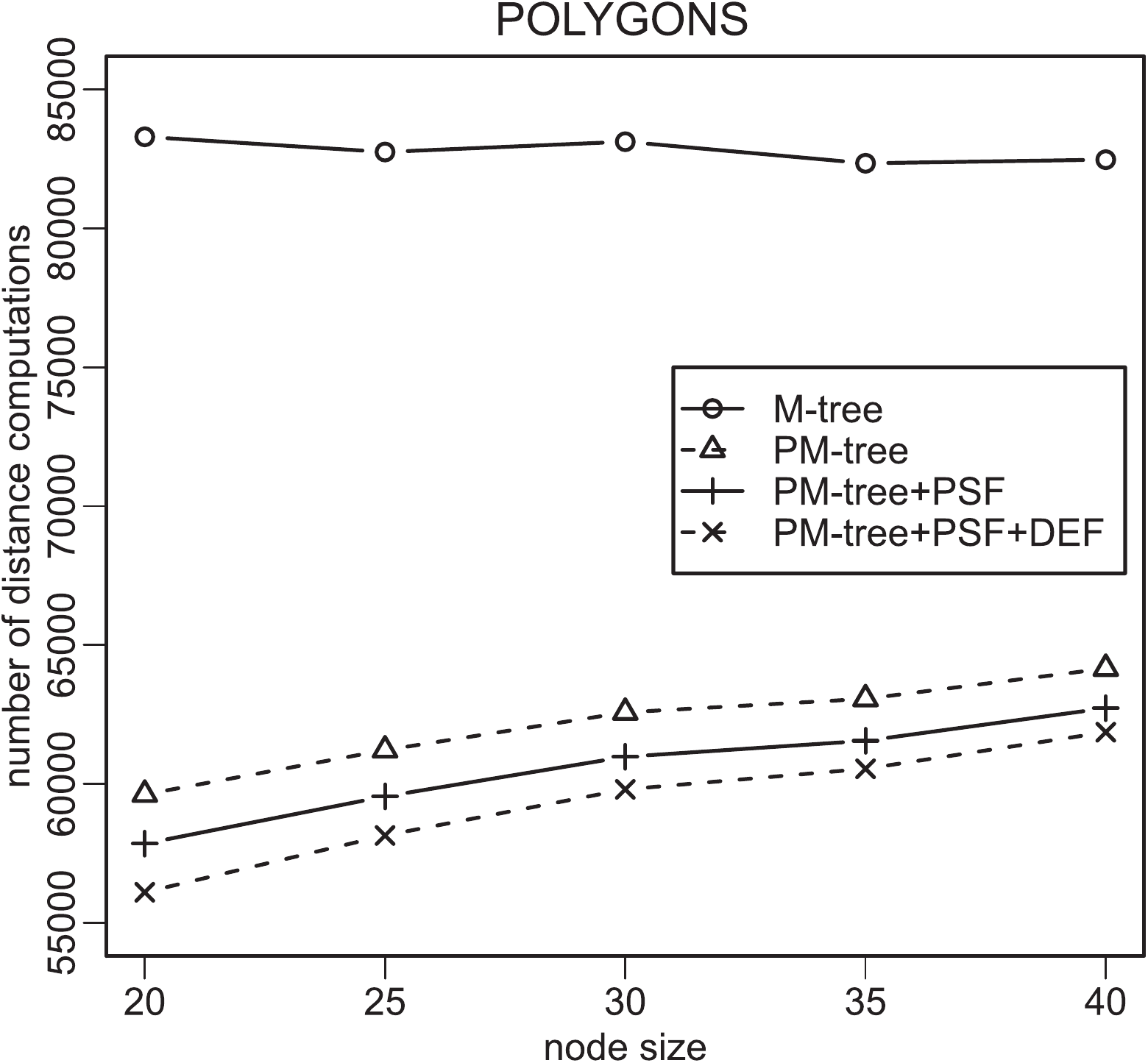}
\caption{Increasing size of (P)M-tree nodes.}
\label{fig_Exp4}
\end{figure}

In the second set of experiments, the impact of (P)M-tree's node size on the distance computations is presented for the Polygons database (see Figure \ref{fig_Exp4}). The performance of M-tree is more or less independent on the node size, while the PM-tree performance slightly decreases with increasing node size.

\begin{figure}[h]
\centering
\includegraphics[width=3.9cm]{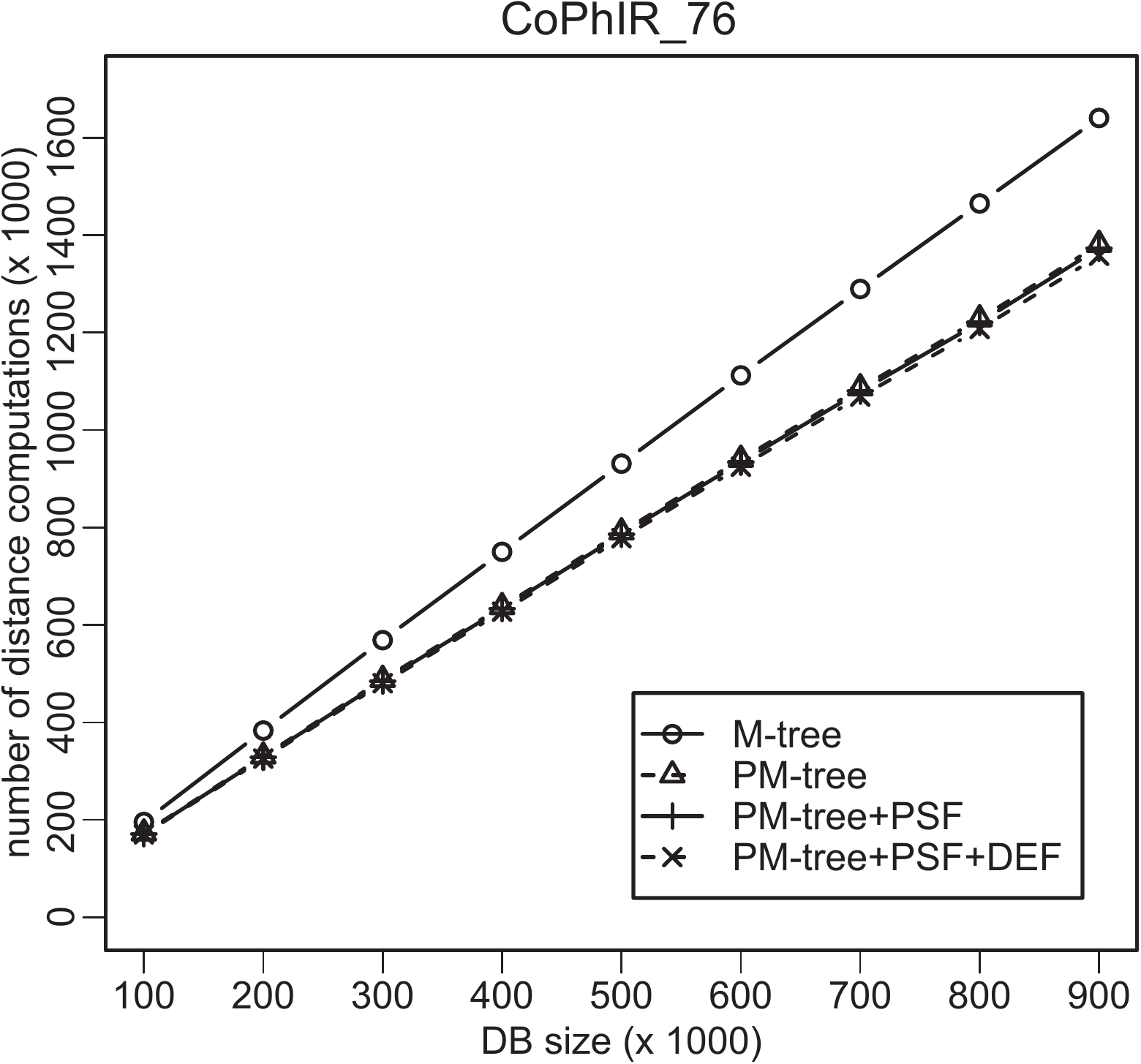}
\includegraphics[width=3.9cm]{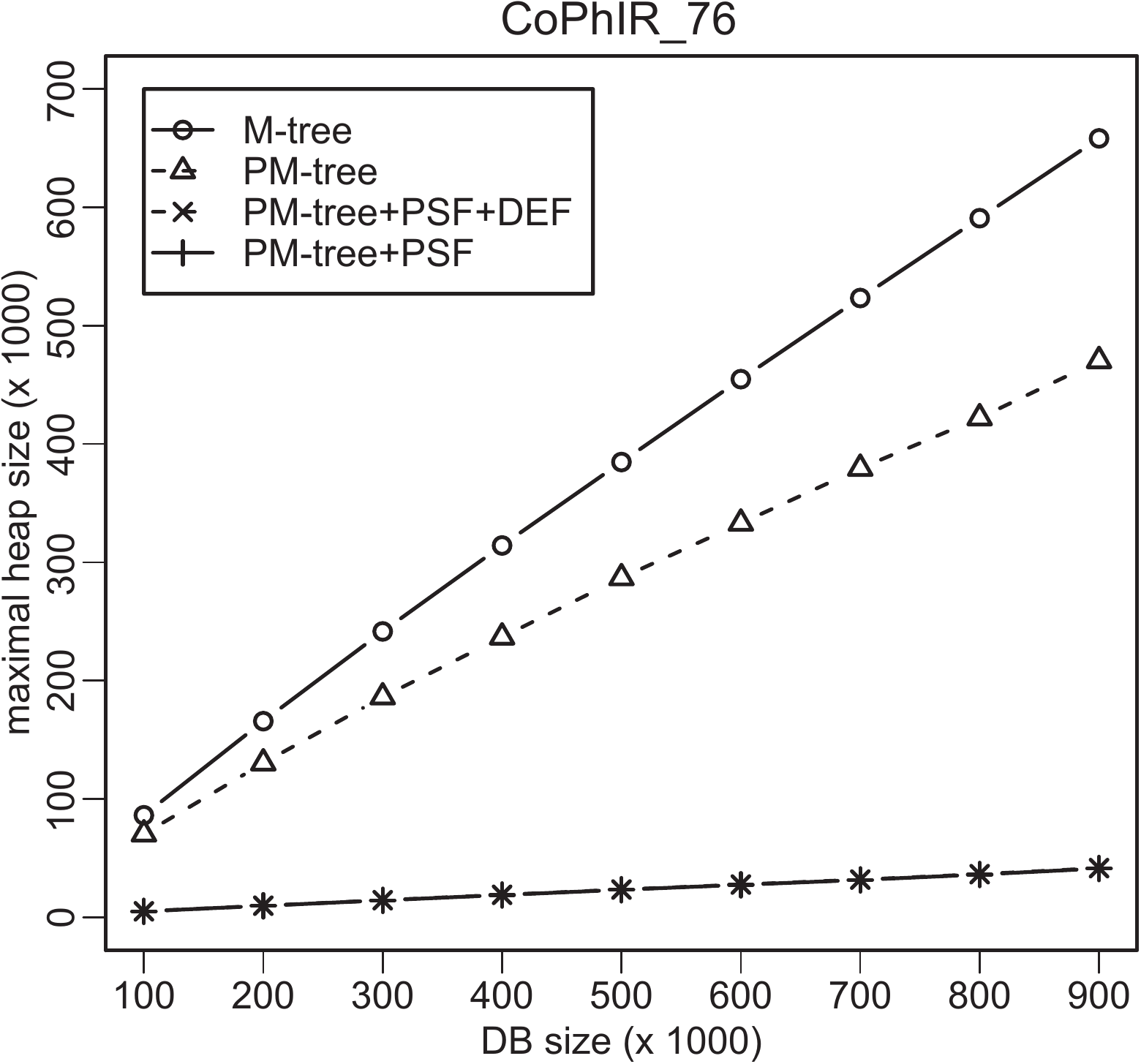}
\includegraphics[width=3.9cm]{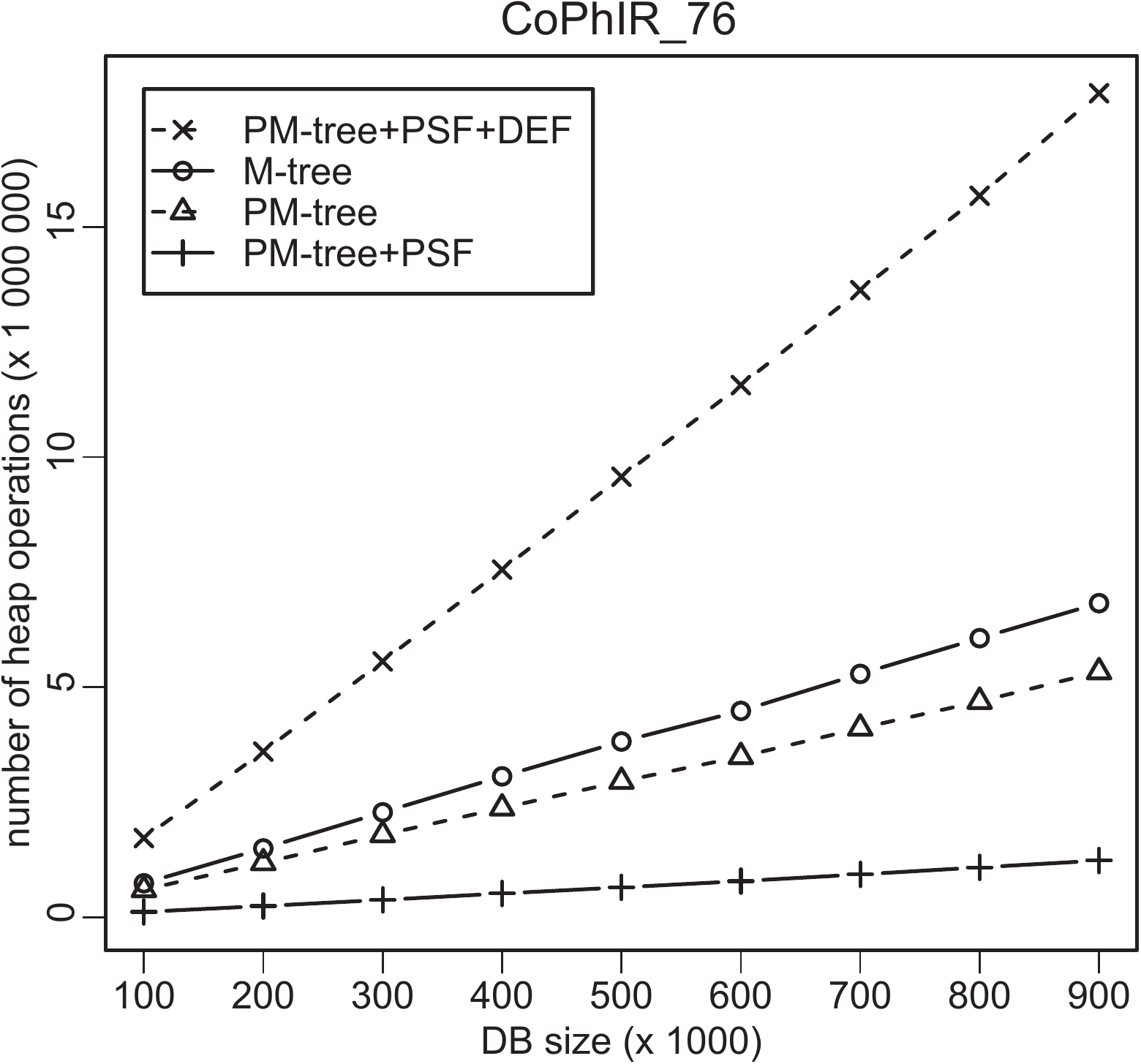}
\caption{Increasing size of Cophir\_76 database: (a) Distance computations (b) Maximal heap size (c) Heap operations}
\label{fig_Exp5}
\end{figure}

The third set of experiments focused on the increasing database size. In Figure \ref{fig_Exp5} the results for Cophir\_76 database are presented. The trend of increasing distance computations is obvious for all MSQ processing methods. However, the situation is dramatically different for the number of heap operations and the maximal heap size, where the PM-tree+PSF beats the M-tree by a factor of 17 in heap operations, and by a factor of 7 in the maximal heap size.  On the other hand, PM-tree+PSF+DEF suffers from a high number of heap operations.

\begin{figure}[h]
\centering
\includegraphics[width=3.9cm]{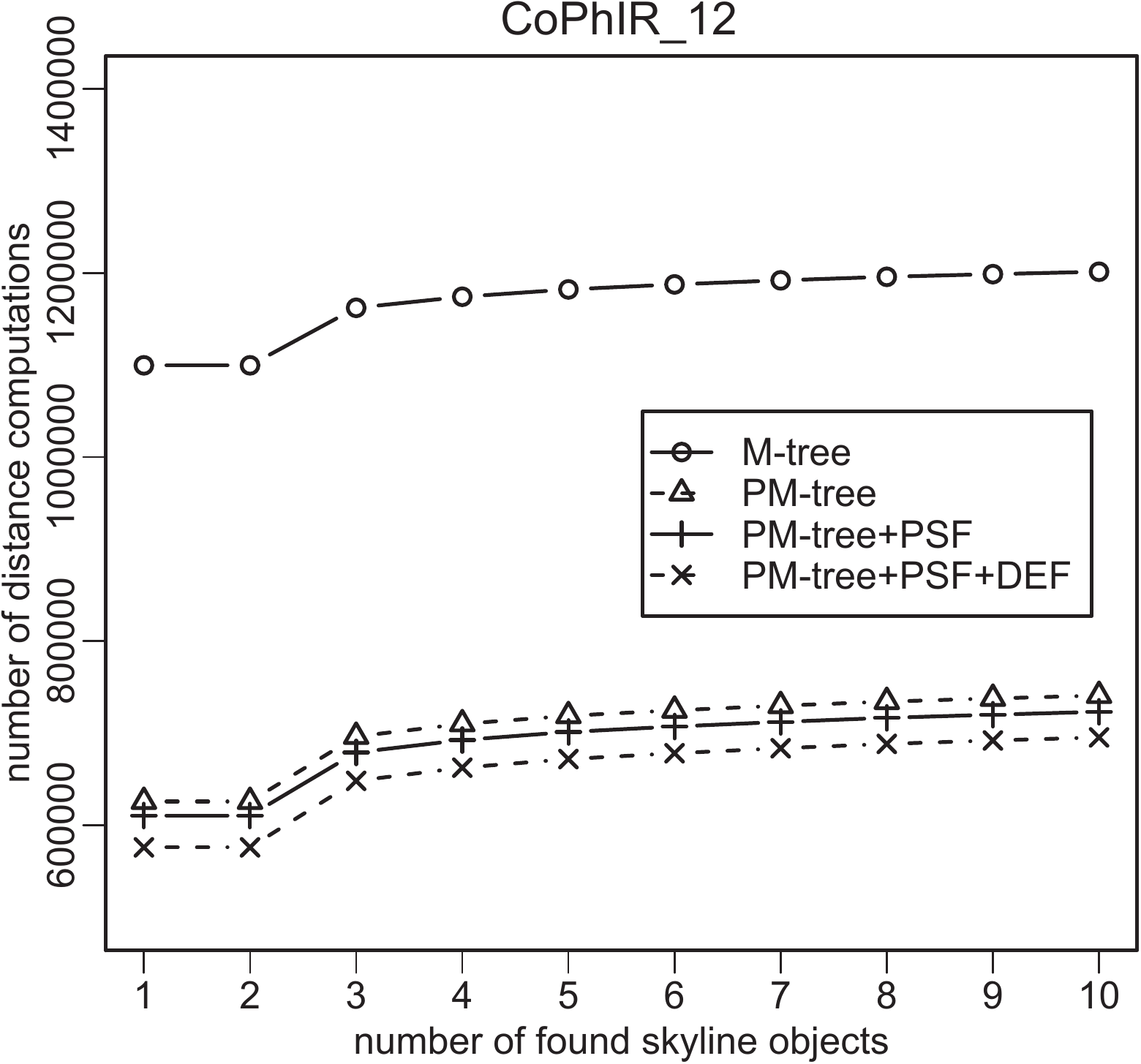}
\includegraphics[width=3.9cm]{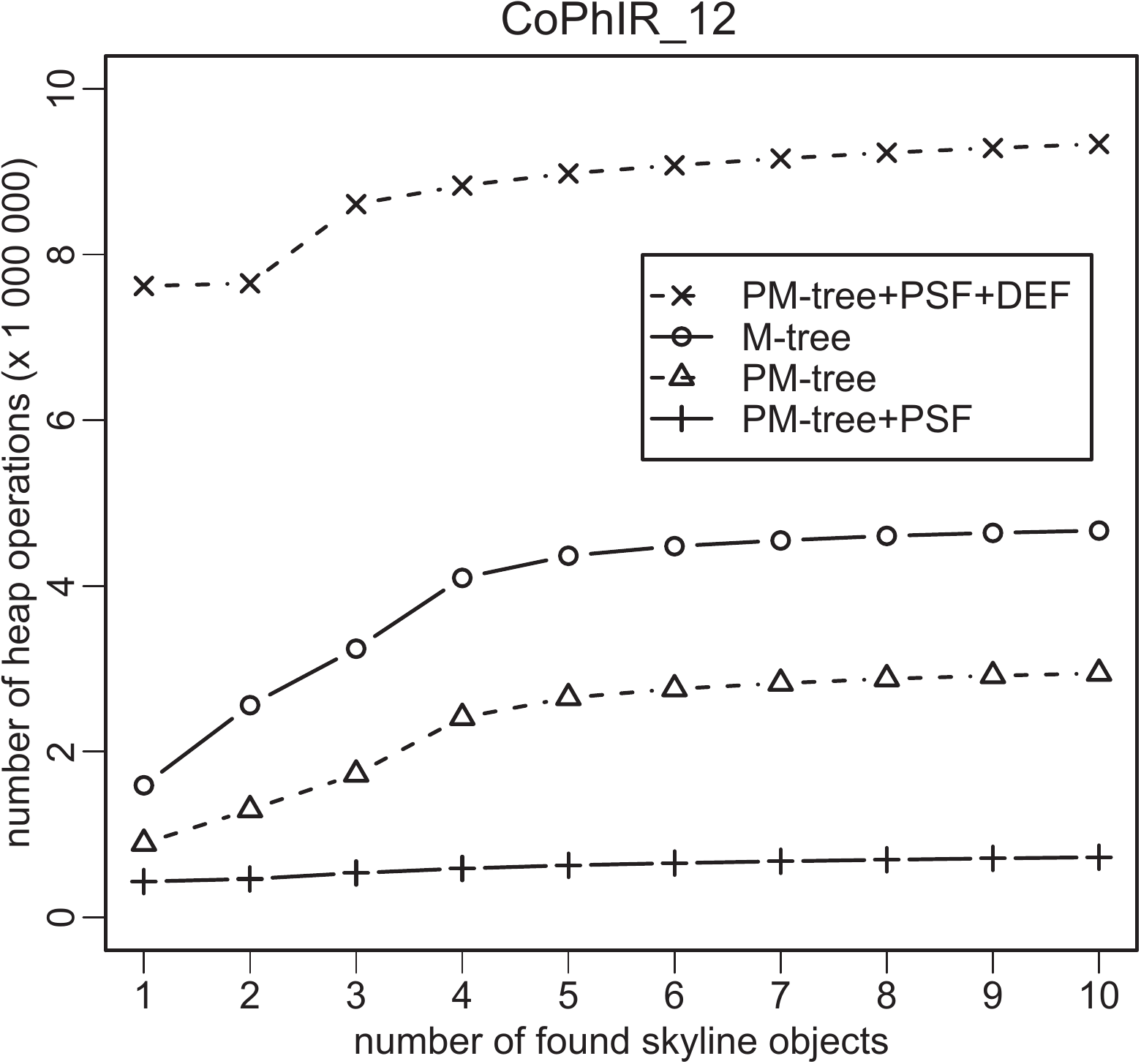}
\caption{Increasing number of retrieved skyline objects: (a) Distance computations (b) Heap operations }
\label{fig_Exp8}
\end{figure}

In the fourth set of experiments, the processing of partial metric skyline queries (as discussed in Section \ref{sec_partial}) is presented, where the results for increasing number of desired skyline objects are presented (see Figure \ref{fig_Exp8}). As mentioned in Section \ref{sec_partial}, the number of distance computations spent on retrieving the first skyline object is almost as expensive as retrieving the entire metric skyline. The situation is slightly better for the number of heap operations, where the M-tree and PM-tree variants are relatively cheaper when retrieving the first skyline object. On the other hand, the costs of PM-tree+PSF are constant and very low (17\% of the M-tree costs).

\begin{figure}[h]
\centering
\includegraphics[width=3.9cm]{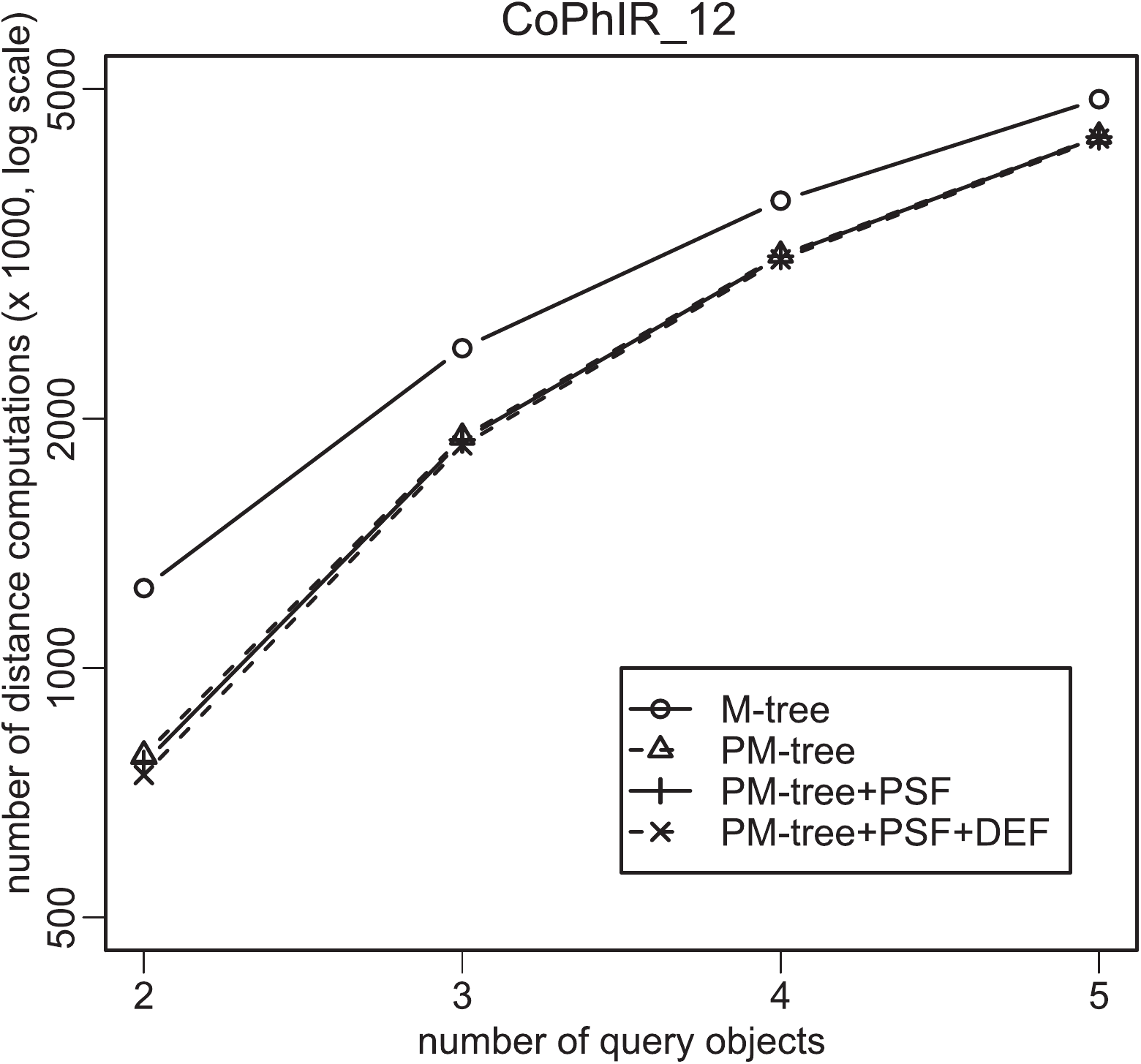}
\includegraphics[width=3.9cm]{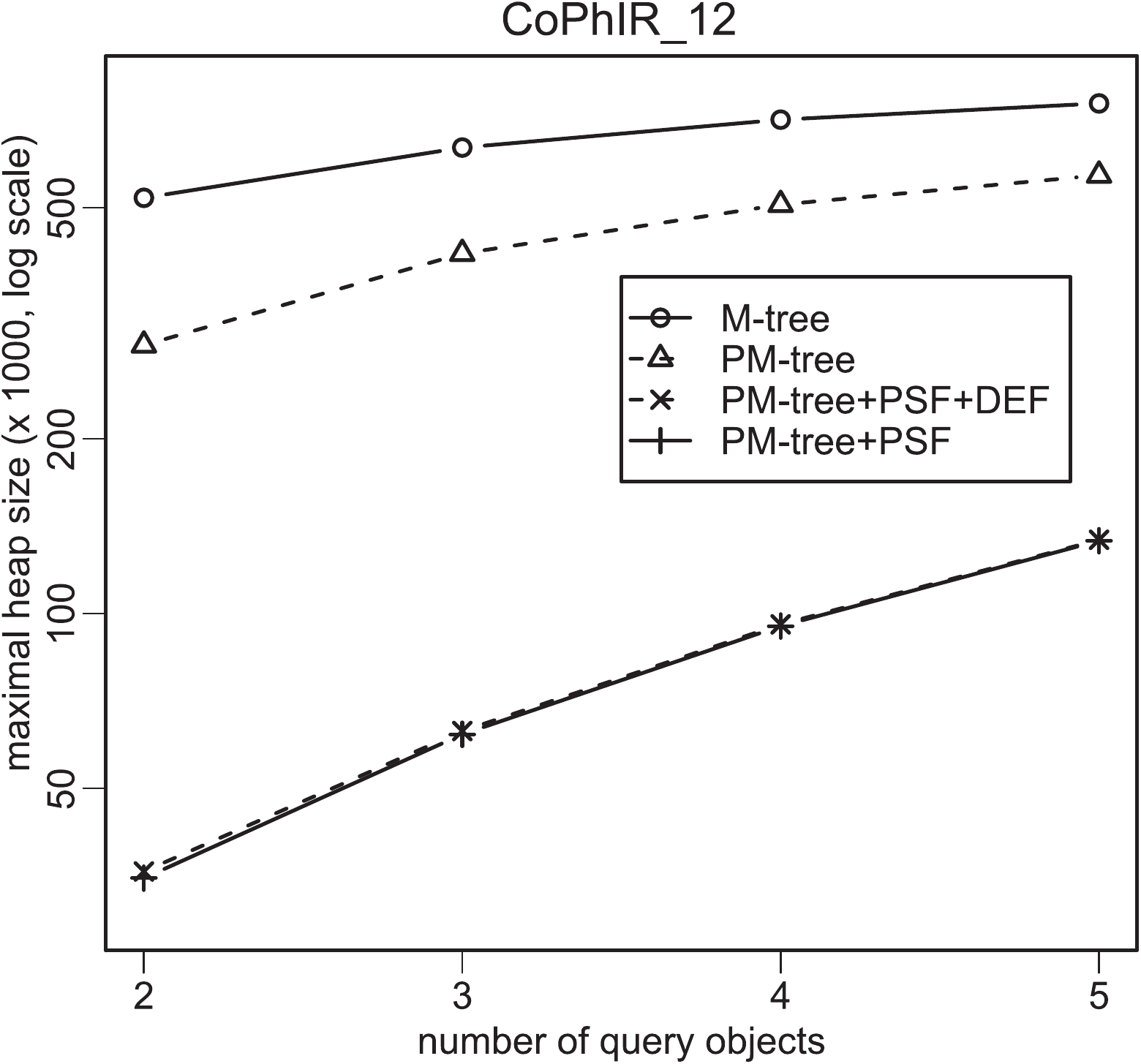}
\includegraphics[width=3.9cm]{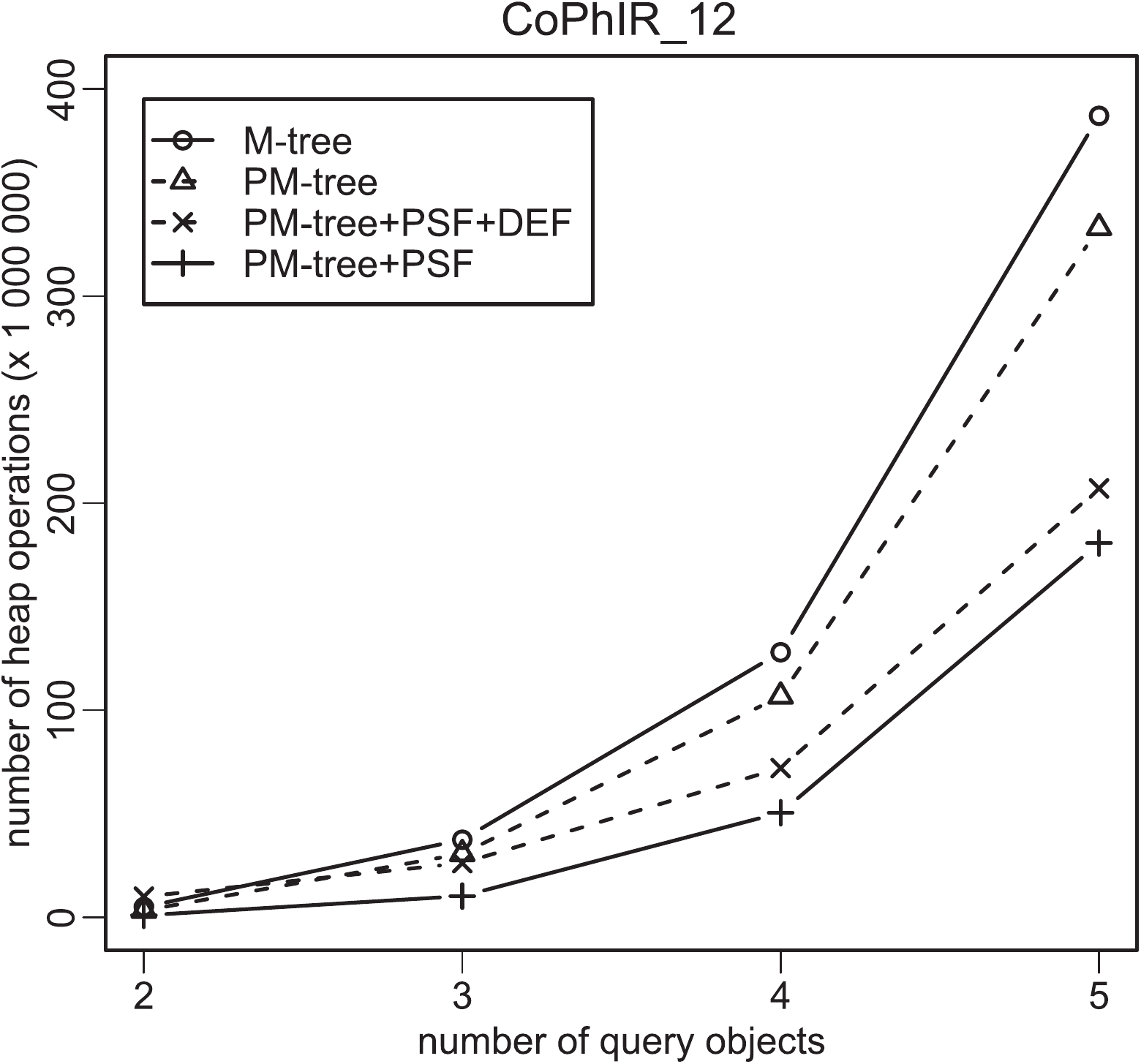}	
\caption{Increasing number of query examples: (a) Distance computations (b) Maximal heap size (c) Heap operations}
\label{fig_Exp6}
\end{figure}

In the fifth set of experiments, the results for increasing number of query examples used in metric skyline queries are presented on the Cophir\_12 database, see Figure \ref{fig_Exp6}. Because the number of skyline objects grows substantially with the increasing number of query examples (retrieving 50, 400, 1750, 4570 skyline objects for 2-, 3-, 4-, and 5-example MSQs), the overall MSQ costs grow substantially as well. Nevertheless, the PM-tree MSQ processing is still much cheaper than the M-tree in the heap size and operations, even for 5 query examples. However, note that for 5 query examples the distance computations of all the methods come close to the costs of simple sequential search.

\begin{figure}[h]
\centering
\includegraphics[width=3.9cm]{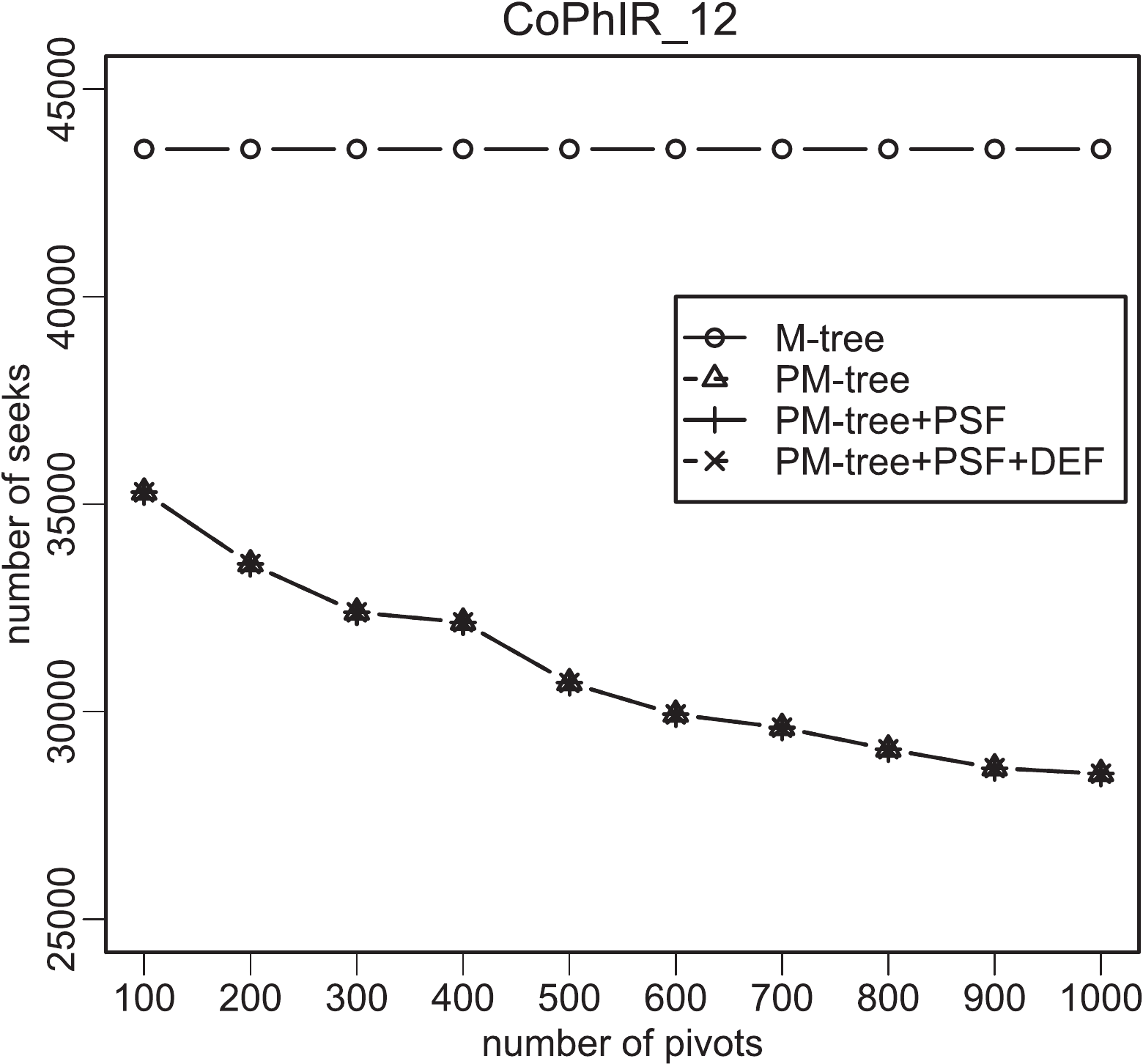}
\includegraphics[width=3.9cm]{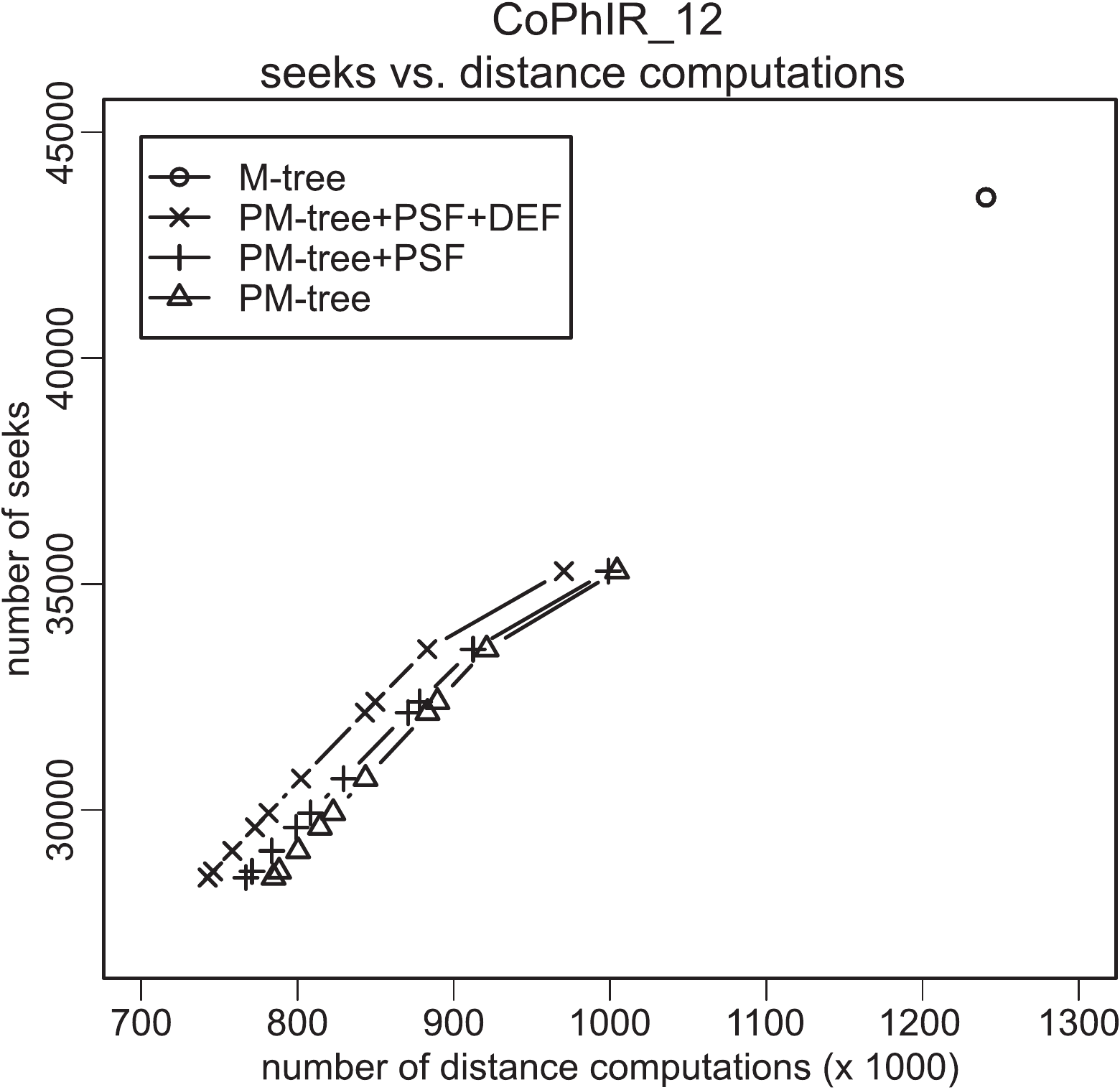}
\caption{Increasing number of pivots: (a) I/O costs (b) I/O costs vs. distance computations}
\label{fig_Exp9}
\end{figure}

Although the I/O costs do not represent a dominant performance component in similarity search\footnote{A single distance computation is generally supposed to be much more expensive than a single I/O operation.}, in the last experiment we present the I/O costs as a supplementary result (CoPhIR\_12, 2 query examples). In particular, in Figure \ref{fig_Exp9}a we give the numbers of logical seeks\footnote{We did not consider any node caching in this experiment.} spent by skyline processing (the seek operation is the most expensive one when fetching a page/PM-tree node from the disk). The PM-tree based MSQ processing spent just 64\% of seek operations required by the M-tree. As for the distance computation costs, also the I/O costs were decreasing with increasing number of pivots.

In Figure \ref{fig_Exp9}b the I/O costs vs. computation costs are shown. As in the first chart, the pairs $\langle$I/O costs, distance computations$\rangle$ were obtained for different numbers of pivots employed by PM-tree. Since the (P)M-tree indices consisted of 79,584 nodes, note that the I/O costs correspond to fetching 55\% of all the index nodes for M-tree and 35\% for PM-tree (1000 pivots). Also note there is linear correlation between the distance computations and I/O costs. 55\%

\subsection{Summary}
The experimentation with M-tree and PM-tree based metric skyline processing has shown that the PM-tree outperforms the M-tree implementation up to 2 times in the number of distance computations, almost 20 times in the number of heap operations and the maximal heap size, and almost 2 times in the I/O costs. The results for maximal heap size are exceptionally important, because a large size of the heap (which is a main-memory structure) would prevent from processing of metric skyline queries on very large databases. 

\section{Conclusions}
In this paper we have proposed a PM-tree based implementation of metric skyline query, a recently proposed multi-example query concept suitable for advanced similarity search in multimedia databases. We have shown that the PM-tree based implementation of metric skylines significantly outperforms the existing M-tree based implementation in all observed costs -- the time, space, and I/O costs. We have also discussed and experimentally evaluated the performance of partial metric skyline processing, where only a limited user-defined number of skyline objects is retrieved. 


\subsection*{Acknowledgments}
This research is partially funded by Czech Science Foundation (GA\v{C}R) Project 201/09/0683.

\bibliographystyle{agsm}
\bibliography{references}  
\end{document}